\documentclass[11pt, nofootinbib]{revtex4-2}

\usepackage{graphicx}
\usepackage{float}
\usepackage{amsmath}
\usepackage{color}
\usepackage{tensor}
\usepackage{epstopdf}
\usepackage{xcolor}
\usepackage{wasysym}
\usepackage{hyperref}
\newcommand{\Eprint}{\href}
\hypersetup{
	colorlinks,
	linkcolor={blue!50!black},
	citecolor={red!50!black},
	urlcolor={blue!80!black}
}

\usepackage{multibib}

\newcommand{\be}{\begin{equation}}
	\newcommand{\ee}{\end{equation}}
\newcommand{\ba}{\begin{eqnarray}}
	\newcommand{\ea}{\end{eqnarray}}

\newcommand{\beq}{\begin{equation}}
	\newcommand{\eeq}{\end{equation}}
\newcommand{\beqa}{\begin{eqnarray}}
	\newcommand{\eeqa}{\end{eqnarray}}
\newcommand{\nn}{\nonumber}
\graphicspath{{plots/}}

\begin{document}

\title[]{Measuring black hole spin through gravitational lensing of pulsars 
}

\author{Amjad Ashoorioon}
\email{amjad@ipm.ir}
\affiliation{School of Physics, Institute for Research in Fundamental Sciences (IPM), P.O. Box 19395-5531, Tehran, Iran}

\author{Mohammad Bagher Jahani Poshteh}
\email{jahani@ipm.ir}
\affiliation{School of Physics, Institute for Research in Fundamental Sciences (IPM), P.O. Box 19395-5531, Tehran, Iran}

\author{Robert B. Mann}
\email{rbmann@uwaterloo.ca}
\affiliation{Department of Physics and Astronomy, University of Waterloo, Waterloo,
	Ontario, N2L 3G1, Canada}
\affiliation{Perimeter Institute for Theoretical Physics, Waterloo, Ontario, N2L
	2Y5, Canada}

\begin{abstract}
We propose a new procedure for measuring the spin of a black hole with an unprecedented accuracy based on the gravitational lensing of millisecond pulsars. We derive the basic equations for lensing by a rotating black hole. We show that the frame dragging effect increases the deflection angle of a light ray co-rotating with the black hole. For the primary (secondary) images the angular positions are larger (smaller) for a rotating black hole by an amount on the order of tens of microarcseconds. The differential time delay of images for the case in which the lens is a rotating black hole is smaller than that in the case of non-rotating lens of the same mass, and it can be larger than a few milliseconds. We show that this quantity offers the possibility of reducing the error of spin measurement to less than one percent if we could measure the differential time delay with accuracy of microseconds. We also study relativistic images that are produced by light rays that rotate around the black hole before reaching the observer.  The angular positions of relativistic images on the same side as the primary (secondary) image are a few microarcseconds larger (smaller) if the black hole is rotating. Furthermore, the differential time delay between relativistic images is about twelve orders of magnitude larger in the case of rotating lens.
\end{abstract}

\maketitle

%
%
%
%
%

\section{Introduction}

The spin of a black hole is an important property that can reveal information about its origin, evolution, and environment. For supermassive black holes,   knowledge of their spin could help us test and constrain different formation scenarios, such as gas cloud collapses, mergers from smaller black holes, or accretion of surrounding matter \cite{Volonteri:2004cf}. The spin also affects the energy output, radiation efficiency, gravitational wave emission, and spacetime distortion of the black hole, which have important implications for feedback processes, observability, and tests of general relativity \cite{Blandford:1977ds,Fabian:2012xr,Reynolds:2019uxi}. Measurement of black hole spin is therefore a crucial task for understanding these fascinating objects and their role in cosmic history.

Many methods have been developed or applied to estimate black hole spin. In the Fe-K$\alpha$ method one observes  the X-ray emission from the iron atoms in the accretion disk, which is distorted by the strong gravity and rotation of its accompanying black hole \cite{Fabian:2000nu,Reynolds:2002np,Miller:2007tj}. Another method  is   X-ray polarimetry, which is based on observing the polarization properties of the X-ray radiation emitted by the accretion disk, because the polarization of X-ray photons is affected by the strong gravity and rotation of the black hole \cite{Dovciak:2008zn,Li:2008zr,Schmoll:2009gq}. Other methods include using the shape and size of the black hole shadow \cite{Bambi:2019tjh,Kawashima:2019ljv,Afrin:2023uzo}, Hawking emission spectrum \cite{Calza:2022ljw}, gravitational wave observations \cite{Vitale:2014mka}, the observed transition of the jet boundary shape \cite{Nokhrina:2019sxv}, observed twisted light propagating near a rotating black hole \cite{Tamburini:2019vrf}, and modelling the jet power and the accretion rate \cite{Nemmen:2019idv,Feng:2017vba}.

In this paper, working in the context of general relativity, we propose a new method for measuring the spin of a black hole with extraordinary accuracy. Our method is based on the gravitational lensing of millisecond pulsars with periods $P\lesssim 30 {\rm ms}$. Millisecond pulsars are very important for astrophysics and cosmology as they can be used as precise clocks to measure various phenomena in the Universe; indeed, pulsars have already been proposed to probe the spacetime around the black holes \cite{Wex:1998wt, Wang:2009yp, Liu:2011ae, Singh:2014nta}. Recently, by observing the tiny changes in the arrival times of the radio pulses from millisecond pulsars, the NANOGrav collaboration has found compelling evidence that  pulsar timing is affected by the gravitational wave background, opening a new window for exploring   gravitational wave emission \cite{NANOGrav:2023hde}. Some evidence for a stochastic gravitational wave background at nanohertz frequencies has also been reported by the Indian and European Pulsar Timing Array collaboration \cite{EPTA:2023akd,EPTA:2023fyk}, and the Chinese Pulsar Timing Array \cite{Xu:2023wog}.

In a gravitational lensing effect, the light rays coming from a source would be deflected as they are passing the black hole \cite{Darwin59,Darwin61}. As a result of this effect, multiple images of the source will be produced near the black hole. A study of this phenomena can help us understand the behavior of gravity in a strong field regime near the black hole.

There have been several attempts to study gravitational lensing by spinning black holes \cite{Bozza:2002af,Bozza:2005tg,Bozza:2006nm,Gyulchev:2006zg,Sereno:2006ss,Werner:2007vu,Bozza:2007gt,Wei:2011nj,Hsieh:2021scb}. Most of them work in either the strong or weak field limits, in which the deflection angle is expanded near the photon circle or near infinity, respectively. However in a gravitational lensing event, the trajectory of the light starts from a source that is usually taken to be at a distance far from the black hole (lens), reaches a turning point near the black hole, and then escapes to the observer, who is also far from the black hole. The light ray therefore experiences both the weak and strong field regimes along its single trajectory, and neither weak nor strong field approximations alone can describe the event accurately. (Another criticism against the strong field approximation has been raised in \cite{Virbhadra:2008ws}.)

Here we generalize the numerical methods of \cite{Virbhadra:1999nm,poshteh2019,Ashoorioon:2022zgu} for non-rotating black holes, to study gravitational lensing by Kerr black holes.  In this method, we numerically solve the lens equation to find the distance of the closest approach (to the black hole) and other lensing quantities.

Different configurations could result in gravitational lensing. In the {\em standard} lensing the black hole (lens) is situated between the observer and the source of light. In this case, we would have two images produced by light rays that are passing the black hole from a far distance and deflecting by an angle less than $\pi$. These images, which are called primary and secondary images (also known as direct images), could have very large magnification.

Along with the primary and secondary images, in standard lensing, we would have two sets of images produced very close to the black hole. These images, which are highly demagnified, are called relativistic images. Relativistic images are produced by light rays that orbit about the black hole, in a radius comparable
but slightly larger than its size, before reaching the observer. It has been shown  \cite{Virbhadra:2008ws} that observation of relativistic images could help us find very accurate values of the mass of the black hole and our distance to it.


The aim of this paper is to study lensing by rotating black holes and comparing the results with those of non-rotating black holes.
In the course of our investigation we shall propose an improved means for determining the black hole spin. In our examples we shall consider the black hole at the Galactic center, Sgr A*, as our lens.

The outline of our paper is as follows. In the next section, we obtain the basic equations for lensing by a Kerr black hole. In Sec.~\ref{sec:ps} we investigate the primary and secondary images produced in lensing by Sgr A*, discussing the advantages of using pulsars as sources. Relativistic images of standard lensing will be studied in Sec.~\ref{sec:rel}. We conclude our paper in Sec.~\ref{sec:con}. Further investigation will be provided in a number of appendixes. We use units in which $G=c=1$ and metric signature +2.

\section{Lensing by Kerr black holes}\label{sec:lensing}

\begin{figure}[htp]
	\centering
	\includegraphics[width=0.5\textwidth]{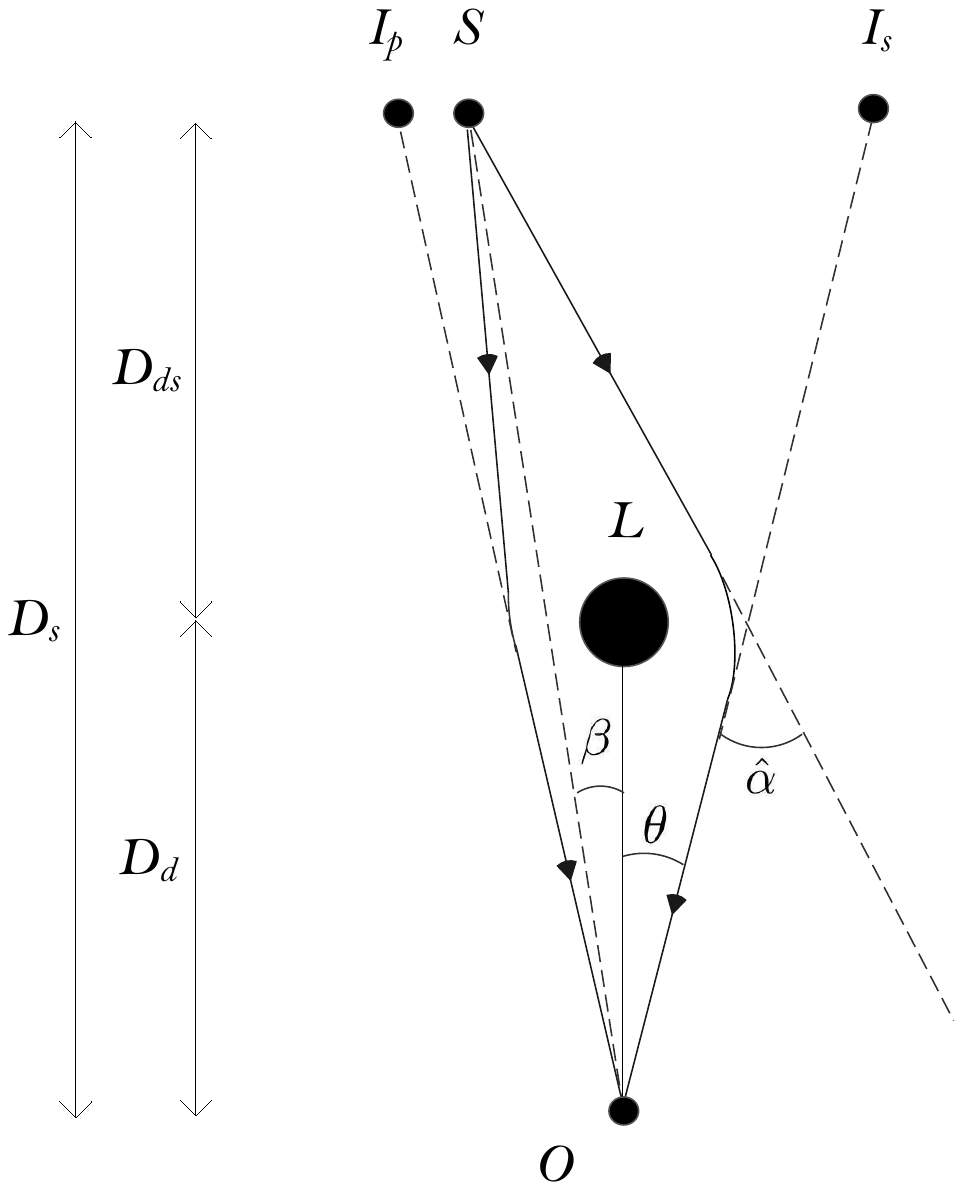}
	\caption{The lens diagram. The light ray  that  is coming from the source $S$ (which we take to be a pulsar) is deflected near the black hole (lens) $L$. The observer $O$  sees images of the source, one on the same side as the source (with respect to the line of sight to $L$), which is called the primary image $I_p$, and another one on the opposite side of the source, which is called the secondary image $I_s$.}
	\label{fig:lens_diag}
\end{figure}

The scenario we study is presented in Figure \ref{fig:lens_diag}. The ray of light that is coming from the source S (which we assume to be a pulsar) is deflected by the black hole. The observer cannot see the source in its actual place but sees its lensed images, namely the primary image (which is on the same side as the source, with respect to the line of sight to the black hole) and the secondary image (which is on the opposite side). Apart from these two images there are two sets of relativistic images that are not shown in Figure \ref{fig:lens_diag}.

Gravitational lensing by non-rotating black holes has been investigated numerically  \cite{Virbhadra:1999nm,poshteh2019,Ashoorioon:2022zgu,Ashoorioon:2021gjs}. Here we use numerical methods to study gravitational lensing by Kerr black holes. The metric of a Kerr spacetime is
\be
ds^2=g_{tt}dt^2+2g_{t\phi}dtd\phi+g_{rr}dr^2+g_{\theta\theta}d\theta^2+g_{\phi\phi}d\phi^2,
\ee
where
\ba
g_{tt}&=&-\left(1-\frac{2mr}{\Sigma}\right),\\
g_{t\phi}&=&\frac{2m a r \sin^2\theta}{\Sigma},\\
g_{rr}&=&\frac{\Sigma}{r^2-2m r+a^2},\\
g_{\theta\theta}&=&\Sigma,\\
g_{\phi\phi}&=&\left(r^2+a^2+\frac{2m a^2 r \sin^2\theta}{\Sigma}\right)\sin^2\theta,
\ea
and
\be
\Sigma\equiv r^2+a^2 \cos^2\theta,
\ee
and $m$ is the mass of Kerr black hole, with $a$  its rotation parameter.

The geodesics are found by using the Lagrangian
\be\label{eqn:lag}
\mathcal{L}=\frac{1}{2}g_{\mu\nu}\dot{x}^\mu\dot{x}^\nu,
\ee
where the dot shows differentiation with respect to some affine parameter. The constants of motion are
\ba
E&=&-\frac{\partial\mathcal{L}}{\partial\dot{t}}=-\left(g_{tt}\dot{t}+g_{t\phi}\dot{\phi}\right), \label{eqn:e}\\
L_z&=&-\frac{\partial\mathcal{L}}{\partial\dot{\phi}}=-\left(g_{t\phi}\dot{t}+g_{\phi\phi}\dot{\phi}\right). \label{eqn:l}
\ea

Consider null geodesics on the equatorial plane of the black hole, for which $\theta=\frac{\pi}{2}$ and $\dot{\theta}=0$ and $\mathcal{L}=0$. (Deviation of light trajectory from equatorial plane will be studied in Appendix \ref{app:dev}.) From~\eqref{eqn:lag} we have
\be\label{eqn:null}
g_{tt}\dot{t}^2+2g_{t\phi}\dot{t}\dot{\phi}+g_{rr}\dot{r}^2+g_{\phi\phi}\dot{\phi}^2=0,
\ee
which results in
\be
\frac{g_{tt}\dot{t}^2+2g_{t\phi}\dot{t}\dot{\phi}}{g_{\phi\phi}\dot{\phi}^2}+\frac{g_{rr}}{g_{\phi\phi}}\left(\frac{dr}{d\phi}\right)^2=-1.
\ee
Substituting for $\dot{t}$ and $\dot{\phi}$ from Eqs.~\eqref{eqn:e} and \eqref{eqn:l} yields
\be\label{eqn:i}
\frac{\left(g_{\phi\phi}E-g_{t\phi}L_z\right)\left(g_{\phi\phi}g_{tt}E+g_{t\phi}g_{tt}L_z-2g_{t\phi}^2E\right)}{g_{\phi\phi}\left(g_{t\phi}E-g_{tt}L_z\right)^2}+\frac{g_{rr}}{g_{\phi\phi}}\left(\frac{dr}{d\phi}\right)^2=-1.
\ee

At the point of closest approach, $r=b$, we have $dr=0$, so
\be\label{eqn:funcofb}
\frac{\left[g_{\phi\phi}(b)E-g_{t\phi}(b)L_z\right]\left[g_{\phi\phi}(b)g_{tt}(b)E+g_{t\phi}(b)g_{tt}(b)L_z-2g_{t\phi}^2(b)E\right]}{g_{\phi\phi}(b)\left[g_{t\phi}(b)E-g_{tt}(b)L_z\right]^2}=-1.
\ee
Equation  \eqref{eqn:funcofb} is a quadratic equation for the impact parameter  $\eta=\frac{L_z}{E}$ in terms  of $b$.  Only the solution
\be\label{eqn:eta}
\eta=\frac{g_{t\phi}(b)-\sqrt{g_{t\phi}^2(b)-g_{\phi\phi}(b)g_{tt}(b)}}{g_{tt}(b)}
\ee
is physical; primary and secondary images do not appear if the plus sign in front of the square root is taken.


By using the parameter $\eta$ we can rewrite Eq.~\eqref{eqn:i} as
\be
\frac{d\phi}{dr}=\sqrt{\frac{g_{rr}}{g_{\phi\phi}}}
\left[\frac{\left(g_{t\phi}\eta - g_{\phi\phi}\right)\left(g_{\phi\phi}g_{tt}+g_{t\phi}g_{tt}\eta-2g_{t\phi}^2\right)}{g_{\phi\phi}\left(g_{t\phi}-g_{tt}\eta\right)^2} - 1\right]^{-1/2}
\equiv\Gamma.
\ee
The deflection angle is then
\be\label{eqn:def}
\hat{\alpha}(b)=2\int_{b}^{\infty}\Gamma \, dr -\pi.
\ee

Eq.~\eqref{eqn:null} can also be written as
\be\label{eqn:ii}
\frac{\left(g_{tt}L_z-g_{t\phi}E\right)\left(g_{\phi\phi}g_{tt}L_z+g_{\phi\phi}g_{t\phi}E-2g_{t\phi}^2L_z\right)}{g_{tt}\left(g_{t\phi}L_z-g_{\phi\phi}E\right)^2}
+\frac{g_{rr}}{g_{tt}}\left(\frac{dr}{dt}\right)^2=-1.
\ee
At $r=b$ we have $dr=0$. This can be used to find the same $\eta$ as in Eq.~\eqref{eqn:eta}. Therefore, from Eq.~\eqref{eqn:ii} we obtain
\be\label{eqn:Pi}
\frac{dt}{dr}=\sqrt{\frac{g_{rr}}{g_{tt}}}
\left[\frac{\left(g_{t\phi}-g_{tt}\eta\right)\left(g_{\phi\phi}g_{tt}\eta+g_{\phi\phi}g_{t\phi}-2g_{t\phi}^2\eta\right)}{g_{tt}\left(g_{t\phi}\eta-g_{\phi\phi}\right)^2}-1\right]^{-1/2}\equiv\Pi,
\ee
from which we compute the time delay
\be
\tau(b)=\int_{b}^{r_s}\Pi\,dr+\int_{b}^{D_d}\Pi\,dr-D_s\sec\beta,
\ee
where $D_s=D_d+D_{ds}$ and $r_s=\sqrt{D_{ds}^2+D_s^2\tan^2\beta}$. Defined in this way, the time delay is  the difference between the time it takes  light to travel from  source to  observer in the physical spacetime (with the black hole), and the time it would take to travel from source to  observer in flat spacetime (with no black hole).

Using   geometrical arguments we can write
\be\label{eqn:eta2}
\eta=D_d\sin\theta.
\ee
and in turn  obtain the Virbhadra-Ellis lens equation \cite{Virbhadra:1999nm}
\be\label{eqn:lens_eq}
\tan\beta=\pm\left\{\tan\theta-\mathcal{D}\left[\tan\theta+\tan\left(\hat{\alpha}-\theta\right)\right]\right\},
\ee
where $\mathcal{D}=D_{ds}/D_s$. The plus (minus) signs correspond to  primary (secondary) images and
\be
\mu=\left(\frac{\sin\beta}{\sin\theta}\frac{d\beta}{d\theta}\right)^{-1}
\ee
is the image  magnification.
To find $\frac{d\beta}{d\theta}$ we use the lens equation \eqref{eqn:lens_eq}, which implies computing $\frac{d\hat{\alpha}}{d\theta}=\frac{d\hat{\alpha}}{db}\frac{db}{d\theta}$. The second factor can be obtained from~\eqref{eqn:eta} and \eqref{eqn:eta2}. We calculate the first factor numerically.

\section{Primary and secondary images}\label{sec:ps}

In this section we investigate the primary and secondary images produced in gravitational lensing by a rotating black hole and compare these to the non-rotating case of the same mass. Specifically, we consider the black hole at the center of the Milky Way galaxy, with mass $M_{{\rm Sgr A^*}}=5.9\times 10^{9} \, {\rm m}\equiv 4.0\times 10^6 M_{\odot}$ and distance $D_d=2.4\times 10^{20} \, {\rm m} \equiv 7.9$ kpc \cite{mnd}.
\footnote{The mass and distance should be known with high precision. To obtain this, one can use relativistic images with the method of \cite{Virbhadra:2008ws}. Relativistic images is the subject of next section.}

\begin{figure}[htp]
	\centering
	\includegraphics[width=0.5\textwidth]{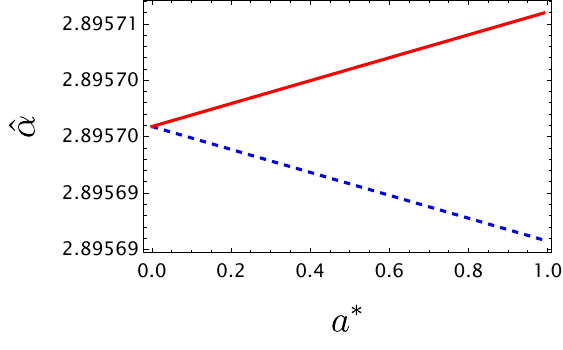}
	\caption{The deflection angle as a function of the rotation parameter for light rays moving in the same direction as the black hole rotation (solid red curve) and for light rays moving in the opposite direction (dashed blue curve). We have taken the black hole at the Galactic center as our lens with mass $m = M_{{\rm Sgr A^*}}=5.9\times 10^{9} \, {\rm m}$ and distance $D_d = 2.4\times 10^{20} \, {\rm m}$. We also consider a fixed value for the distance of closest approach $b = 1.68 \times 10^{15}  \, {\rm m}$. $\hat{\alpha}$ is in {\em arcseconds} and $a^*=a/m$ is dimensionless.}
	\label{fig:defang_vs_a}
\end{figure}

In Figure \ref{fig:defang_vs_a} we see that, for a fixed value of closest approach to the black hole $b$, the deflection angle has a small but clear change with increasing rotation parameter. Here we introduce the dimensionless rotation parameter $a^*=a/m$. In the solid red curve we have considered the light rays that move in the same direction as the rotation of the black hole (in other words the  momentum of the light ray is parallel to the momentum of an observer co-rotating with the black hole). In the dashed blue curve we have considered the light rays that move in the opposite direction as the rotation of the black hole. We see that if a light ray moves in the same direction as the black hole rotation the deflection angle would be larger than that in the static case. This is due to the frame dragging in the rotating black hole spacetime. On the other hand, for a light ray that moves in the opposite direction the deflection angle would be smaller than that in the static case.

\begingroup
\begin{table*}
	\caption{
		Image positions $\theta$ and deflection angles $\hat{\alpha}$ of direct images due to lensing by rotating (ro) as well as non-rotating (nr) black holes. $\beta$ is the source angular position and the subscripts $p$ and $s$ refer to primary and secondary images, respectively. $\beta$ is in {\em milliarcseconds} while $\theta$ and $\hat{\alpha}$ are in {\em arcseconds}. We have used $m = M_{{\rm Sgr A^*}}=5.9\times 10^{9} \, {\rm m}$, $D_d = 2.4\times 10^{20} \, {\rm m}$ and $\mathcal{D}=0.5$. For the rotating black hole we have taken the rotation parameter to be $a^*=0.99$.
	}\label{tab:ps_ang}
	\begin{ruledtabular}
		\begin{tabular}{l cccc cccc}
			\multicolumn{1}{c}{$\beta$}&
			\multicolumn{4}{c}{Rotating black hole}&
			\multicolumn{4}{c}{Non-rotating black hole}\\
			&$\theta_{p,{\rm ro}} $&$\hat{\alpha}_{p,{\rm ro}}$&$\theta_{s,{\rm ro}}$&$\hat{\alpha}_{s,{\rm ro}}$&$\theta_{p,{\rm nr}}$&$\hat{\alpha}_{p,{\rm nr}}$&$\theta_{s,{\rm nr}}$&$\hat{\alpha}_{s,{\rm nr}}$\\
			\hline
			$0	 	 $&$  1.4462896   $&$	2.8925792    		$&$  1.4462845  $&$    	  2.8925690     $&$  1.4462850     $&$	  2.8925701   	  $&$ 1.4462850    $&$     2.8925701        $\\
			$1	 	 $&$  1.4467902	  $&$   2.8915804    		$&$  1.4457847  $&$	   	  2.8935695     $&$  1.4467857	   $&$    2.8915713   	  $&$ 1.4457873	   $&$	   2.8935745        $\\
			$2	 	 $&$  1.4472909	  $&$   2.8905817    		$&$  1.4452850  $&$		  2.8945700     $&$  1.4472863	   $&$    2.8905726   	  $&$ 1.4452875    $&$	   2.8945750        $\\
			$3	 	 $&$  1.4477915	  $&$   2.8895830    		$&$  1.4447853  $&$    	  2.8955705     $&$  1.4477869	   $&$    2.8895739   	  $&$ 1.4447878    $&$	   2.8955755        $\\
			$4	 	 $&$  1.4482921	  $&$   2.8885842    		$&$  1.4442855  $&$    	  2.8965710     $&$  1.4482876	   $&$    2.8885752   	  $&$ 1.4442880    $&$     2.8965760        $\\
			$5	 	 $&$  1.4487928	  $&$   2.8875855    		$&$  1.4437858  $&$		  2.8975715     $&$  1.4487882	   $&$    2.8875764   	  $&$ 1.4437883    $&$	   2.8975765        $\\
			$6	 	 $&$  1.4492934	  $&$   2.8865868   		$&$  1.4432860  $&$    	  2.8985720     $&$  1.4492889	   $&$    2.8865777  	  $&$ 1.4432885	   $&$     2.8985771        $\\
		\end{tabular}
	\end{ruledtabular}
\end{table*}
\endgroup

\begingroup
\begin{table*}
	\caption{
		Magnification $\mu$ and (differential) time delay $\tau$ ($t_d = \tau_s-\tau_p$) of direct images due to lensing by rotating (ro) as well as non-rotating (nr) black holes. $\beta$ and the subscripts $p$ and $s$ are as in Table \ref{tab:ps_ang}. (Differential) time delays are in {\em seconds}. $m$, $D_d$, $\mathcal{D}$, and $a^*$ are as in Table \ref{tab:ps_ang}.
	}\label{tab:ps_time}
	\begin{ruledtabular}
		\begin{tabular}{l cccc cccc}
			\multicolumn{1}{c}{$\beta$}&
			\multicolumn{4}{c}{Rotating black hole}&
			\multicolumn{4}{c}{Non-rotating black hole}\\
			&$\mu_{p,{\rm ro}} $&$\tau_{p,{\rm ro}}$&$\mu_{s,{\rm ro}}$&$t_{d,{\rm ro}}$&$\mu_{p,{\rm nr}}$&$\tau_{p,{\rm nr}}$&$\mu_{s,{\rm nr}}$&$t_{d,{\rm nr}}$\\
			\hline
			$0	 	 $&$  \times  	  $&$	988.82422   	$&$  \times	     $&$    	  -0.0005390    $&$  \times    	   $&$	  988.82417 	  $&$ \times  	    $&$    0      		    $\\
			$1		 $&$  724.31347	  $&$   988.76974   	$&$  -722.52789  $&$	   	  0.1083618     $&$  724.31554	   $&$    988.76968 	  $&$ -722.52663    $&$	   0.1086857        $\\
			$2		 $&$  362.28205	  $&$   988.71529 		$&$  -361.13907  $&$		  0.2172626     $&$  362.28309	   $&$    988.71524 	  $&$ -361.13844    $&$	   0.2175866        $\\
			$3	 	 $&$  241.60491	  $&$   988.66089 		$&$  -240.67613  $&$    	  0.3261634     $&$  241.60560	   $&$    988.66084 	  $&$ -240.67571    $&$	   0.3264876        $\\
			$4	 	 $&$  181.26634	  $&$   988.60652 		$&$  -180.44466  $&$    	  0.4350643     $&$  181.26686	   $&$    988.60647 	  $&$ -180.44435    $&$    0.4353886        $\\
			$5	 	 $&$  145.06320	  $&$   988.55219 		$&$  -144.30578  $&$		  0.5439652     $&$  145.06362	   $&$    988.55214 	  $&$ -144.30553    $&$	   0.5442896        $\\
			$6	 	 $&$  120.92777	  $&$   988.49790 		$&$  -120.21319  $&$    	  0.6528662     $&$  120.92812	   $&$    988.49785 	  $&$ -120.21298    $&$    0.6531907        $\\
		\end{tabular}
	\end{ruledtabular}
\end{table*}
\endgroup

In Tables \ref{tab:ps_ang} and \ref{tab:ps_time} we  present the image positions, deflection angles, magnification, and (differential) time delay of primary and secondary images for different values of angular source positions. For illustrative purposes, here we have taken the black hole to be halfway between the source and the observer (see also Appendix \ref{app:ps}). We neglect the mass of stars in the Galactic nucleus in comparison to the mass of the supermassive black hole and assume the spacetime around the black hole to be vacuum. We have considered the black hole to be non-rotating, as well as rotating with rotation parameter $a^*=0.99$. We assume that the momentum of the light ray producing the primary image is  in the same direction as the rotation of the black hole; referring to Figure \ref{fig:lens_diag}, this means that the black hole rotates counter-clockwise. Note that the image angle is relative to the lens-observer axis in Figure \ref{fig:lens_diag} and is always positive.

From Table \ref{tab:ps_ang} we see for a non-rotating black hole that, for angular source position $\beta = 0$, the deflection angles of the primary and secondary images are the same; this is due to its spherical symmetry. This symmetry is broken for the rotating black hole. We see that, even if $\beta = 0$, there is a difference between the deflection angles of the primary and secondary images. We have assumed that the black hole is rotating in the same direction as the light ray producing the primary image, which means that  $L_z$ (or equivalently the rotation parameter $a$) is negative when calculating the lensing parameters of the secondary image (see \cite{Iyer:2009wa,Barlow:2017uty,Beachley:2018tux}). Therefore, for $\beta = 0$, the deflection angle of the primary image is larger due to the frame dragging effect.

\begin{figure}[htp]
	\centering
	\includegraphics[width=0.48\textwidth]{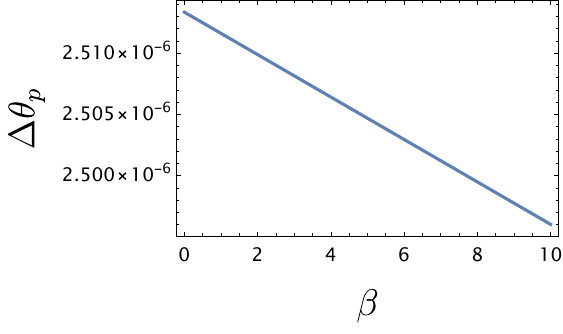}
	\includegraphics[width=0.48\textwidth]{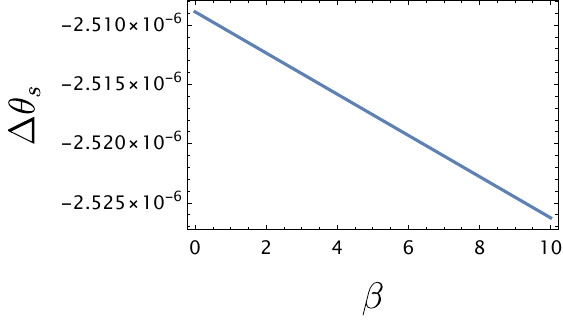}
	\caption{\textit{Left}: The difference between the angular positions of primary image in the case of rotating black hole with  $a^*=0.99$ and the case of non-rotating black hole, $\Delta\theta_p=\theta_{p,a^* = 0.99}-\theta_{p,a^* = 0}$. \textit{Right}: The corresponding value for the case of secondary image,  $\Delta\theta_s=\theta_{s,a^* = 0.99}-\theta_{s,a^*=0}$. We have taken $m$, $D_d$, and $\mathcal{D}$ as in Table \ref{tab:ps_ang}. $\Delta\theta$ is in {\em arcseconds} and $\beta$ is in {\em milliarcseconds}.
	}
	\label{fig:Delta_theta}
\end{figure}

We also see from Table \ref{tab:ps_ang} that for primary images the deflection angle is larger by parts in $10^5$ -- $10^6$ for rotating black holes. However for secondary images the deflection angle is smaller for a rotating black hole. We observe that the angular position of primary images increases with increasing angular source position $\beta$. On the other hand, the angular position of secondary images decreases with increasing $\beta$. This holds for both rotating and static black holes.  We also see that the angular positions of primary (secondary) images are larger (smaller) if the black hole is rotating; for the values considered in this table, the difference can be as large as tens of microarcseconds. The difference between angular position of primary images for a rotating black hole --- with $a^*=0.99$ --- and a non-rotating black hole, $\Delta\theta_p=\theta_{p,a^* = 0.99}-\theta_{p,a^* = 0}$, decreases as  the angular source position increases. The corresponding value $\Delta\theta_s=\theta_{s,a^* = 0.99}-\theta_{s,a^*=0}$ for the secondary image is negative and its absolute value increases with increasing $\beta$. This behaviour is  depicted in Figure \ref{fig:Delta_theta}.

\begin{figure}[htp]
	\centering
	\includegraphics[width=0.48\textwidth]{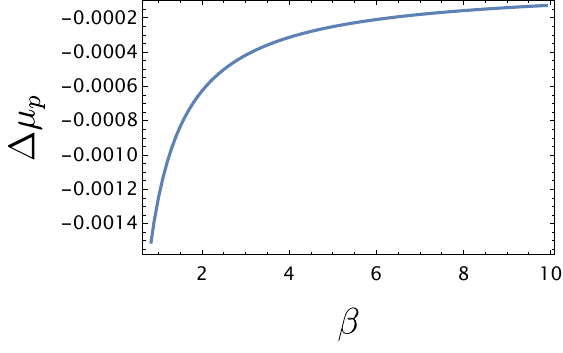}
	\includegraphics[width=0.48\textwidth]{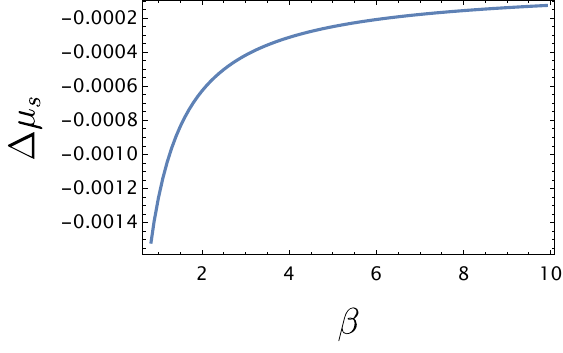}
	\caption{\textit{Left}: The difference  $\Delta\mu_p=\mu_{p,a^* = 0.99}-\mu_{p,a^* = 0}$ between the magnification of primary images for rotating ($a^*=0.99$) and  non-rotating black holes.  \textit{Right}: The corresponding value for   secondary images, $\Delta\mu_s=\mu_{s,a^* = 0.99}-\mu_{s,a^*=0}$. We have taken $m$, $D_d$, and $\mathcal{D}$ as in Table \ref{tab:ps_ang}. $\beta$ is in {\em milliarcseconds}.
	}
	\label{fig:Delta_mu}
\end{figure}

From Table \ref{tab:ps_time} we find, for both the rotating and non-rotating cases, that the absolute value of the magnification of primary and secondary images decreases for increasing  angular position of the source. For primary images the magnification is larger if the black hole is non-rotating. The difference between the magnification of the primary image in the rotating and non-rotating cases, $\Delta\mu_p=\mu_{p,a^* = 0.99}-\mu_{p,a^* = 0}$, is a negative quantity and its absolute value decreases with  increasing   angular position of the source. This can be seen in the left panel of Figure \ref{fig:Delta_mu}. Furthermore, the absolute value of the magnification of secondary images is larger if the lens is  rotating. Since the magnification of the secondary image is defined to be negative, the quantity $\Delta\mu_s=\mu_{s,a^* = 0.99}-\mu_{s,a^* = 0}$ is negative and its  absolute value decreases as $\beta$ increases, as shown in  the right panel of Figure \ref{fig:Delta_mu}.

\begin{figure}[htp]
	\centering
	\includegraphics[width=0.5\textwidth]{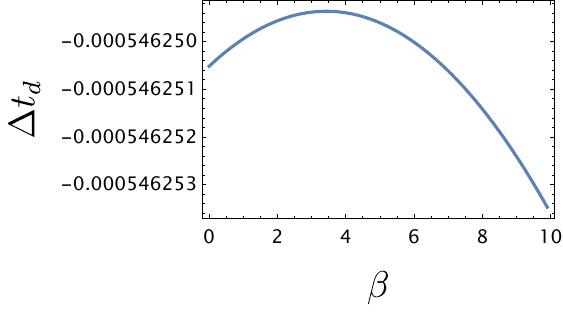}
	\caption{The difference between the differential time delay for a rotating black hole and a non-rotating black hole, $\Delta t_d=t_{d,a^*=0.99}-t_{d,a^*=0}$. We have taken $m$, $D_d$, and $\mathcal{D}$ as in Table \ref{tab:ps_ang}. $\Delta t_d$ is in {\em seconds} and $\beta$ is in {\em milliarcseconds}.
	}
	\label{fig:Delta_t_d}
\end{figure}

In Table \ref{tab:ps_time} we also see that the time delay of primary images is larger for a rotating black hole, but in both cases decreases as the angular position of the source increases. We have not represented the time delay of secondary images. Instead we have shown the differential time delay $t_d=\tau_s-\tau_p$. This quantity is of more observational importance, because, if the source is pulsating, every phase in its period would appear in the secondary image, $t_d$ seconds after appearing in the primary image. For $\beta = 0$, the differential time delay is zero for the static black hole as one expects from  symmetry considerations. However for a rotating black hole (with $\beta = 0$) this quantity has a negative value, which means that, due to the frame dragging effect, the time delay for the primary images is larger than that of secondary images. The differential time delay increases as the angular position of the source increases, and is larger if the black hole is  non-rotating. For the values chosen in Table \ref{tab:ps_time}, and for large values of $\beta$ the difference can be as large as a few milliseconds. The difference between the differential time delay between a rotating black hole and a non-rotating black hole, $\Delta t_d=t_{d,a^*=0.99}-t_{d,a^*=0}$, is a negative quantity and its absolute value first gets smaller and then starts to increase with $\beta$, as depicted in Figure \ref{fig:Delta_t_d}.

In standard lensing we do not see the source in its actual place and therefore we cannot measure $\beta$ directly. However, it is possible to find $\beta$ from the angular positions of direct images. Suppose that we observe primary and secondary images; these can be distinguished from their magnifications. From  Table \ref{tab:ps_time} we see that the absolute value of the magnification of the primary image is larger than that of the secondary image  for both the rotating and non-rotating cases. We can find the distance to the source by, for example, using the red shift of the images. Suppose also that we know the mass of the lens and our distance to it. What we do not know are $\beta$ and if the lens is rotating or not. The primary image can be due to lensing by a rotating black hole and a source at position $\beta_{p,{\rm ro}}$ or due to lensing by a non-rotating black hole and a source at position $\beta_{p,{\rm nr}}$. On the other hand, the secondary image can be due to a rotating black hole and a source at position $\beta_{s,{\rm ro}}$ or due to lensing by a non-rotating black hole and a source at position $\beta_{s,{\rm nr}}$. Since the source and lens are the same in both cases, either $\beta_{p,{\rm ro}}=\beta_{s,{\rm ro}}$ or $\beta_{p,{\rm nr}}=\beta_{s,{\rm nr}}$ is correct. In this way it is possible to find $\beta$ and determine if the black hole is rotating.  Note that here we have assumed that the black hole is either static or rotating in the same direction as the momentum  of the light ray coming from the primary image. A more general method to find $\beta$, $a^*$, and the sense of rotation will be presented in Appendix \ref{app:rev}.

Most significantly, a measurement of differential time delay could help us find the rotation parameter with unprecedented precision. In fact, for Sgr A*, a measurement of differential time delay with accuracy of  microseconds would result in a measurement of the spin parameter with less than 1 percent error. For comparison,  the supermassive black hole M87* has $M_{{\rm M87^*}}= 6.5\times 10^9 M_{\odot}$ and distance $D_d= 16.8$ Mpc \cite{Akiyama:2019eap};  such accuracy in spin measurement would be achieved if only we could measure the differential time delay with accuracy of $10^{-3}$ seconds.

We note that a precision of 1 percent is better than the most accurate estimates of supermassive black hole spin that we can find, i.e.~\cite{Tamburini:2019vrf,Feng:2017vba}, which are based on twisted light and accretion-jet analysis. Compared to other methods,  X-ray reflection is one of the most accurate measurements of black hole spin via  spectroscopy of the supermassive black hole Mrk 1044,  with an error of about $2 \%$ \cite{Mallick:2018bdl}. Gravitational wave observations from the LIGO-Virgo network yield measurements of  spin   with error of about $5 \%$ \cite{Vitale:2014mka}.   X-ray polarimetry spin measurements have errors of $\mathcal{O}(10) \%$ \cite{Mikusincova:2023bon}.

The dilemma is that to achieve precision of one percent in the rotation parameter, using the method presented in this paper, one needs to know $\beta$ very accurately, which in turn requires a precise knowledge of $\theta$. In fact, to obtain the rotation parameter with an accuracy of one percent, one needs to measure the image position with an accuracy of nanoarcseconds.

It is conceivable to reach such precision in the near future.
The time of arrival of radio pulses from millisecond pulsars can be measured extremely precise. Therefore, their period can be estimated with a very high precision --- of the order of a femtosecond \cite{Berthereau:2023aod,Guo:2021bqa,Serylak:2022kna,Corongiu:2023gft}. We conclude that, if we have a pulsar as our source behind the black hole, a measurement of differential time delay with accuracy of the order {$10^{-6}$} seconds  indeed seems feasible.
\footnote{\ {In practice one should look for similar patterns in the light curve of pulsars obtain from primary and secondary images.}}


\section{Relativistic images}\label{sec:rel}

If a light ray passes a black hole very closely (with the distance of closest approach of the same order as the horizon radius), it could orbit about the black hole before approaching the observer. The images thus produced are called relativistic images and appear very close to the lens. Relativistic images are known to be very demagnified and so are very hard to  observe. In this section we study first order relativistic images that are produced by light rays that rotate the black hole once before reaching the observer (of course there are higher order relativistic images produced by light rays that orbit more around the lens --- see Appendix \ref{app:2nd}). As before, we  assume that the momentum of the light ray coming from the primary image is in the same direction as the rotation of the black hole.

In Tables \ref{tab:1st_ang} and \ref{tab:1st_time} we have presented the lensing parameters of first order relativistic images for both rotating and non-rotating black holes. In the latter case, $\theta_{1s,{\rm nr}}\simeq\theta_{1p,{\rm nr}}$ -- the angular position of  a first order relativistic image on the same side as the secondary image is approximately the same as the angular position of a first order relativistic image on the same side as the primary image.  However we see from Table \ref{tab:1st_ang} that for a rotating black hole the angular position of a first order relativistic image on the same side as the primary image is larger than that of an image on the same side as the secondary image. In fact, although it is not obvious from the numbers in Table \ref{tab:1st_ang}, the deflection angle $\hat{\alpha}_{1p,{\rm ro}}$ is about $10^{-5}$ arcseconds larger than $\hat{\alpha}_{1s,{\rm ro}}$. The deflection angles and therefore the angular positions of first order images are nearly insensitive to the angular position of the source $\beta$; changes are of the order of $10^{-12}$ microarcseconds, and so are not shown in the table. As can be seen in Table \ref{tab:1st_ang}, for first order relativistic images on the same side as the primary image, the angular position is larger if the black hole is rotating. On the other hand, for first order relativistic images on the same side as the secondary image, the angular position is smaller if the black hole is rotating.

We see from Table \ref{tab:1st_time} that the magnifications of relativistic images are much smaller than  unity. The magnification of an image produced on the same side as the secondary image is defined to be negative and,  for non-rotating black holes, its absolute value is approximately the same as the magnification of an image produced on the same side as the primary image,  i.e.~$\mu_{1s,{\rm nr}}\simeq-\mu_{1p,{\rm nr}}$.  For rotating black holes, the magnification of an image on the primary side is about one order of magnitude smaller than the absolute value of the magnification of an image on the secondary side:  frame dragging has made the images look smaller. The absolute values of the magnifications of first order relativistic images decrease with increasing  angular position $\beta$ of the source.
Furthermore,
the magnifications of images on the primary side are larger for non-rotating black holes,  whereas for rotating black holes the absolute values of the magnifications of images on the secondary side is larger.  We can also see in table \ref{tab:1st_time} that the difference between the magnifications of first order relativistic images for rotating  and non-rotating  black holes decreases with increasing $\beta$.

\begingroup
\begin{table}
	\scriptsize
	\caption{
		Image positions $\theta$ and deflection angle $\hat{\alpha}$ of first order relativistic images due to lensing by rotating (ro) as well as non-rotating (nr) black holes. The subscripts $1p$ and $1s$ refer to the first order relativistic images produced on the same side as the primary and secondary images, respectively. All angles are in {\em microarcseconds}. $m$, $D_d$, $\mathcal{D}$, and $a$ are as in Table \ref{tab:ps_ang}.
The values are the same for all $\beta$ of the order of microarcseconds.
	}\label{tab:1st_ang}
	\begin{ruledtabular}
		\begin{tabular}{l cccc cccc}
			\multicolumn{1}{c}{$\beta$}&
			\multicolumn{4}{c}{Rotating black hole}&
			\multicolumn{4}{c}{Non-rotating black hole}\\
			&$\theta_{1p,{\rm ro}} $&$\hat{\alpha}_{1p,{\rm ro}}$&$\theta_{1s,{\rm ro}}$&$\hat{\alpha}_{1s,{\rm ro}}$&$\theta_{1p,{\rm nr}}$&$\hat{\alpha}_{1p,{\rm nr}}$&$\theta_{1s,{\rm nr}}$&$\hat{\alpha}_{1s,{\rm nr}}$\\
			\hline
			$0-6	 	 $&$  35.416579   $&$1.2960012 \times 10^{12}$&$ 13.427240 $&$1.2960012 \times 10^{12}$&$  26.381137       $&$1.2960012 \times 10^{12}$&$26.381137$&$1.2960012 \times 10^{12}$\\
		\end{tabular}
	\end{ruledtabular}
\end{table}
\endgroup

\begingroup
\begin{table*}
	\caption{
		Magnifications $\mu$, time delay $\tau$, and differential time delay $t_d=\tau_{1s}-\tau_{1p}$ of first order relativistic images due to lensing by rotating (ro) as well as non-rotating (nr) black holes. The subscripts $1p$ and $1s$ refer to the first order relativistic images produced on the same side as the primary and secondary images, respectively. $\beta$ is in {\em microarcseconds} and (differential) time delays are in {\em seconds}. $m$, $D_d$, $\mathcal{D}$, and $a$ are as in Table \ref{tab:ps_ang}.
	}\label{tab:1st_time}
	\begin{ruledtabular}
		\begin{tabular}{l cccc cccc}
			\multicolumn{1}{c}{$\beta$}&
			\multicolumn{4}{c}{Rotating black hole}&
			\multicolumn{4}{c}{Non-rotating black hole}\\
			&$\mu_{1p,{\rm ro}}$&$\tau_{1p,{\rm ro}}$&$\mu_{1s,{\rm ro}}$&$t_{d,{\rm ro}}$&$\mu_{1p,{\rm nr}}$&$\tau_{1p,{\rm nr}}$&$\mu_{1s,{\rm nr}}$&$t_{d,{\rm nr}}$\\
			\hline
			$0$&$	   \times    	    $&$  2742.309  $&$  \times   		    $&$  -444.222     $&$	  	  \times   	       $&$ 2543.837    $&$  	\times     	  $&$     0     	   $\\
			$1$&$ 2.8 \times 10^{-12}   $&$  2742.309  $&$  -6.2 \times 10^{-11}$&$  -444.222     $&$    8.5 \times 10^{-12}   $&$ 2543.837    $&$-8.5 \times 10^{-12}$&$1.98\times 10^{-9}$\\
			$2$&$ 1.4 \times 10^{-12}   $&$  2742.309  $&$  -3.1 \times 10^{-11}$&$  -444.222     $&$    4.2 \times 10^{-12}   $&$ 2543.837    $&$-4.2 \times 10^{-12}$&$3.96\times 10^{-9}$\\
			$3$&$ 9.5 \times 10^{-13}   $&$  2742.309  $&$  -2.1 \times 10^{-11}$&$  -444.222     $&$    2.8 \times 10^{-12}   $&$ 2543.837    $&$-2.8 \times 10^{-12}$&$5.93\times 10^{-9}$\\
			$4$&$ 7.1 \times 10^{-13}   $&$  2742.309  $&$  -1.6 \times 10^{-11}$&$  -444.222     $&$    2.1 \times 10^{-12}   $&$ 2543.837    $&$-2.1 \times 10^{-12}$&$7.90\times 10^{-9}$\\
			$5$&$ 5.7 \times 10^{-13}   $&$  2742.309  $&$  -1.2 \times 10^{-11}$&$  -444.222     $&$    1.7 \times 10^{-12}   $&$ 2543.837    $&$-1.7 \times 10^{-12}$&$9.88\times 10^{-9}$\\
			$6$&$ 4.7 \times 10^{-13}   $&$  2742.309  $&$  -1.0 \times 10^{-11}$&$  -444.222     $&$    1.3 \times 10^{-12}   $&$ 2543.837    $&$-1.3 \times 10^{-12}$&$1.19\times 10^{-8}$\\
		\end{tabular}
	\end{ruledtabular}
\end{table*}
\endgroup

The time delay associated with the first order relativistic images are nearly insensitive to the angular position of the source --- in fact, for both rotating and non-rotating black holes, time delay of first order relativistic images on the same side as the primary image decreases very slowly as $\beta$ gets larger. It is obvious from Table \ref{tab:1st_time} that the time delay  of an image on the same side as the primary image is larger if the lens is rotating.

We do not present the time delay of first order relativistic images on the same side as the secondary image. Instead, in Table \ref{tab:1st_time}, we have presented the differential time delay for both the rotating and non-rotating cases.  For a rotating black hole the differential time delay is negative, which means that the time delay of first order relativistic images on the primary side is larger for these black holes. The absolute value of the differential time delay of first order images is much larger if the lens is rotating. The differential time delay of first order images for non-rotating black hole increase slightly with $\beta$. However, for rotating black holes the absolute value of the differential time delay decreases slightly with increasing $\beta$.


\section{Concluding remarks}\label{sec:con}

Frame dragging is very effective in the vicinity of rotating black hole.  We have found that differential time delays of relativistic images produced by such black hole lenses can be as much as $10^{12}$ times larger than their counterparts in the static case. Once relativistic images are observed, these differential time delays could be used  to find out if the black hole is rotating, and for retrolensing can provide an improved means for determining its spin.


The difference between the angular positions of primary and secondary images for the case of a rotating black hole and non-rotating black hole with the same mass parameter is of the order of microarcseconds,  one order of magnitude smaller than the resolution of today's observational facilities, such as the Event Horizon Telescope \cite{Fish:2016jil,Akiyama:2017rcc}. However the differential time delay between primary and secondary images can be about a millisecond smaller for a rotating lens.

Along with primary and secondary images, in standard lensing we have two sets of relativistic images produced by light rays that rotate the black hole very closely before continuing their path to the observer. These images are highly demagnified;  however the magnification increases when the configuration is close to perfect alignment in which the source, lens, and the observer are on the same line. Angular positions of relativistic images for   lensing due to rotating black holes deviate from those of static black holes  by a few microarcseconds.  The absolute value of the differential time delay associated with relativistic images is of the order of $10^{12}$ times larger than the differential time delay of relativistic images in static black hole lensing, offering interesting new prospects for determining the rotation of
a black hole, as noted above.




For future work, further extensions that take acceleration and/or electric charge into account would be of interest.  A key question will be what sets of parameters result in degeneracies. We also emphasize that the theory describing the black hole spacetime in this paper is general relativity. Alternative theories of gravity will yield different predictions for both the angular positions of images and their differential time delays \cite{poshteh2019}.


\medskip

{\emph{{Acknowledgments:}}}
This work was supported in part by the Natural Sciences and Engineering Research Council of Canada. MBJP would like to thank CERN for their hospitality during part of this work.

\appendix

\section{Direct images with different $\mathcal{D}$ and rotation parameter}\label{app:ps}

\begin{figure}[htp]
	\centering
	\includegraphics[width=0.5\textwidth]{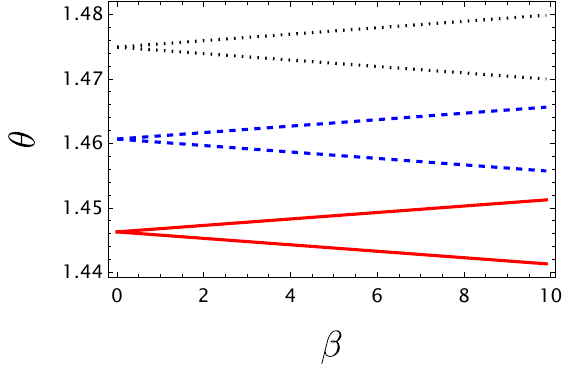}
	\caption{Primary and secondary image positions as a function of angular source position. The solid red curve corresponds to $\mathcal{D}=0.5$,  the curve to  $\mathcal{D}=0.51$, and   the dotted black one to $\mathcal{D}=0.52$. Lines with negative slope represent the secondary image position and  lines with positive slope correspond to the primary image position. We have taken $m$ and $D_d$ as in Table \ref{tab:ps_ang} and $a^*=0.99$; $\theta$ is in {\em arcseconds} and $\beta$ is in {\em milliarcseconds}.  Note that the values of the primary and secondary image positions do not exactly overlap for   $\beta = 0$; as seen in Table \ref{tab:ps_ang} there is a difference of the order of $10^{-6}$ {\em arcseconds} between them.
	}
	\label{fig:theta_diff_D}
\end{figure}

In Sec.~\ref{sec:ps} we have considered direct images, i.e.~primary and secondary images, with $\mathcal{D}=0.5$ in which case the black hole is halfway between observer and source. However, the positions of the images, their magnifications, and their (differential) time delay, depend on $\mathcal{D}$. In Figure \ref{fig:theta_diff_D} we have plotted direct image positions for different values of $\mathcal{D}$, fixing the value of $D_d$ to be the distance to the Sgr A* black hole. Hence different values of $\mathcal{D}$ correspond to different values of the distance from the  source to the lens. For a rotating black hole considered in Figure \ref{fig:theta_diff_D} we see that, for a fixed value of $\beta$, the angular position of both primary and secondary images get larger as we increase $\mathcal{D}$.

\begin{figure}[htp]
	\centering
	\includegraphics[width=0.48\textwidth]{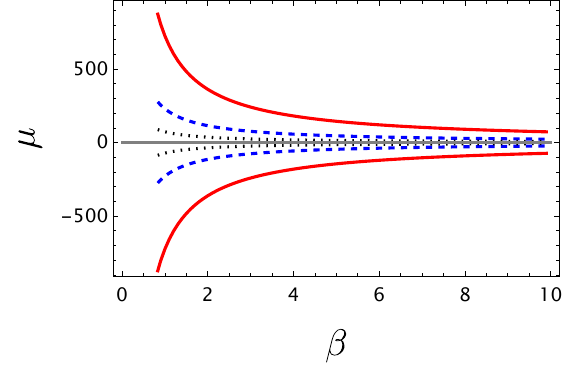}
	\includegraphics[width=0.48\textwidth]{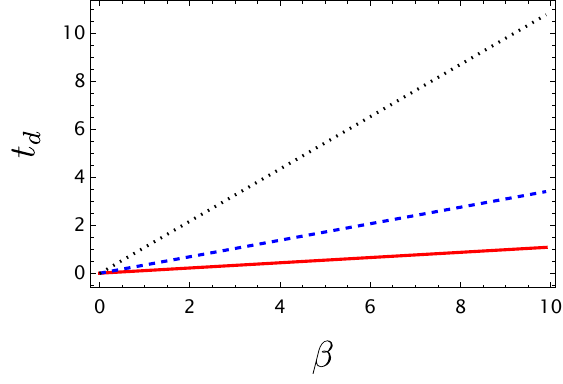}
	\caption{\textit{Left}: Magnification of primary and secondary images. The magnification is positive for primary images and negative for secondary images \textit{Right}: Differential time delay, in units of {\em seconds}, as a function of angular source position. In the solid red plots we have $\mathcal{D}=0.5$, in the dashed blue plots we set $\mathcal{D}=0.05$, and for the dotted black plots we take $\mathcal{D}=0.005$. We have taken $m$ and $D_d$ as in Table \ref{tab:ps_ang} and $a^*=0.99$. $\beta$ is in {\em milliarcseconds}.
	}
	\label{fig:mu_t_d_diff_D}
\end{figure}

In the left panel of Figure \ref{fig:mu_t_d_diff_D} the magnifications of primary and secondary images are presented for different values of $\mathcal{D}$.  We find that, for a fixed value of the angular source position, the absolute value of the magnification is larger for larger values of $\mathcal{D}$. In the right panel of Figure \ref{fig:mu_t_d_diff_D} the differential time delays of primary and secondary images are plotted for different values of $\mathcal{D}$. We find that, for a fixed value of $\beta$, the differential time delay gets larger as $\mathcal{D}$ gets smaller. The slope of the plot of differential time delay versus $\beta$ is also larger for smaller values of $\mathcal{D}$.

\begin{figure}[htp]
	\centering
	\includegraphics[width=0.48\textwidth]{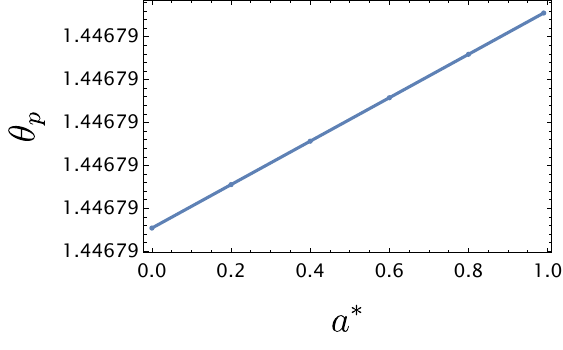}
	\includegraphics[width=0.48\textwidth]{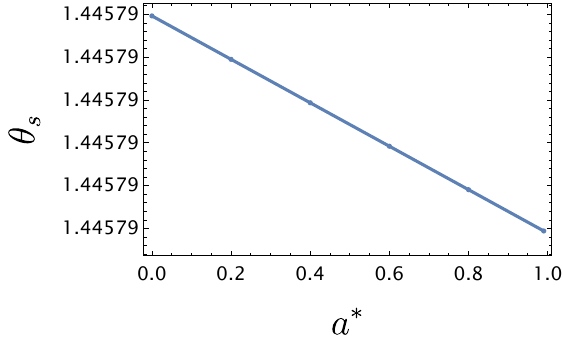}
	\caption{\textit{Left}: Angular position of the primary image as a function of rotation parameter. \textit{Right}: Angular position of the secondary image as a function of rotation parameter. We have taken $m$ and $D_d$ as in Table \ref{tab:ps_ang} and $\beta = 1$ {\em milliarcsecond}. Image positions are in units of {\em arcseconds}.
	}
	\label{fig:theta_vs_a}
\end{figure}

\begin{figure}[htp]
	\centering
	\includegraphics[width=0.48\textwidth]{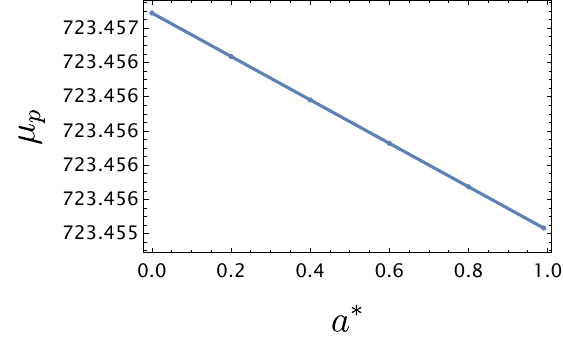}
	\includegraphics[width=0.48\textwidth]{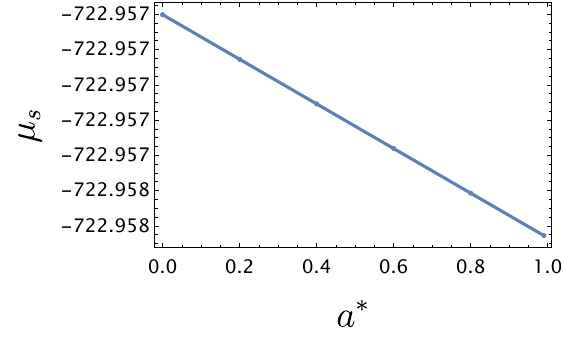}
	\caption{\textit{Left}: Magnification of the primary image as a function of rotation parameter. \textit{Right}: Magnification of the secondary image as a function of rotation parameter. We have taken $m$ and $D_d$ as in Table \ref{tab:ps_ang} and $\beta = 1 $ {\em milliarcsecond}.
	}
	\label{fig:mu_vs_a}
\end{figure}

\begin{figure}[htp]
	\centering
	\includegraphics[width=0.5\textwidth]{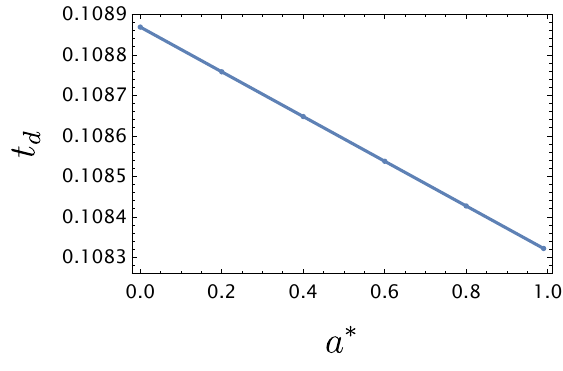}
	\caption{Differential time delay, in units of {\em seconds}, as a function of rotation parameter. We have taken $m$ and $D_d$ as in Table \ref{tab:ps_ang} and $\beta = 1 $ {\em milliarcsecond}.
	}
	\label{fig:t_d_vs_a}
\end{figure}

To see how the image position, magnification, and differential time delay change by changing the rotation parameter, we have presented several plots of lensing parameters versus the rotation parameter in Figures \ref{fig:theta_vs_a}, \ref{fig:mu_vs_a}, and \ref{fig:t_d_vs_a}.  In Figure \ref{fig:theta_vs_a} we see that the angular positions of primary (secondary) images increase (decrease) with increasing rotation parameter. From Figure \ref{fig:mu_vs_a} it is obvious that the magnification of the primary image decreases as the rotation parameter gets larger and the absolute value of the magnification of secondary image increases with the rotation parameter. In Figure \ref{fig:t_d_vs_a} we see that the differential time delay decreases with increasing rotation parameter.

\section{Deviation from the equatorial plane}\label{app:dev}

Let us assume that the trajectory of the light ray comming from the source deviate from the equatorial plane of the Kerr black hole by an angle $\delta$, i.e.~$\theta = \pi/2 + \delta$. By using Eqs.~\eqref{eqn:lag}-\eqref{eqn:l}, we can write
\be\label{eqn:b1}
\frac{\left(g_{\phi\phi}-g_{t\phi}\bar{\eta}\right)\left(g_{\phi\phi}g_{tt}+g_{t\phi}g_{tt}\bar{\eta}-2g_{t\phi}^2\right)}{g_{\phi\phi}\left(g_{t\phi}-g_{tt}\bar{\eta}\right)^2}+\frac{g_{rr}}{g_{\phi\phi}}\left(\frac{dr}{d\phi}\right)^2+\frac{g_{\theta\theta}}{g_{\phi\phi}}\left(\frac{d\delta}{d\phi}\right)^2=-1,
\ee
in which the impact parameter is given by
\be
\bar{\eta} = \frac{g_{tt}g_{t\phi}g_{\theta\theta}\left(\frac{d\delta}{d\phi}\right)^2-g_{t\phi}^3+g_{tt}g_{t\phi}g_{\phi\phi}+\zeta^{1/2}}{g_{tt}^2g_{\theta\theta}\left(\frac{d\delta}{d\phi}\right)^2-g_{tt}g_{t\phi}^2+g_{tt}^2g_{\phi\phi}},
\ee
where
\ba
\zeta &=& g_{t\phi}^6-g_{tt}g_{t\phi}^4g_{\theta\theta}\left(\frac{d\delta}{d\phi}\right)^2-3g_{tt}g_{t\phi}^4g_{\phi\phi}+2g_{tt}^2 g_{t\phi}^2 g_{\phi\phi}g_{\theta\theta}\left(\frac{d\delta}{d\phi}\right)^2+3 g_{tt}^2g_{t\phi}^2g_{\phi\phi}^2 -g_{tt}^3g_{\phi\phi}^2g_{\theta\theta}\left(\frac{d\delta}{d\phi}\right)^2\nn\\
&-&g_{tt}^3g_{\phi\phi}^3.
\ea
In the definition of impact parameter $\bar{\eta}$ we have set $r = b$.

Now we solve Eq.~\eqref{eqn:b1} for $\frac{d\phi}{dr}$ and perturb the result around $\theta=\pi/2$ by the angle $\delta$. Assuming both $\delta$ and its variation to be small, we find to first order approximation in $\delta$ the same equation for the deflection angle as in Eq.~\eqref{eqn:def}. We also note that the Virbhadra-Ellis lens equation stays the same if the light ray deviate from the equatorial plane of the black hole. Therefore, we conclude that deviation from $\theta = \pi/2$ plane by a small angle does not affect the images' position in gravitational lensing.

Using Eqs.~\eqref{eqn:lag}-\eqref{eqn:l}, we can also write
\be\label{eqn:b4}
\frac{\left(g_{tt}\tilde{\eta}-g_{t\phi}\right)\left(g_{t\phi}g_{\phi\phi}-2g_{t\phi}^2\tilde{\eta}+g_{tt}g_{\phi\phi}\tilde{\eta}\right)}{g_{tt}\left(g_{\phi\phi}-g_{t\phi}\tilde{\eta}\right)^2}+\frac{g_{rr}}{g_{tt}}\left(\frac{dr}{dt}\right)^2+\frac{g_{\theta\theta}}{g_{tt}}\left(\frac{d\delta}{dt}\right)^2=-1,
\ee
in which
\be
\tilde{\eta} = \frac{g_{t\phi}^3-g_{tt}g_{t\phi}g_{\phi\phi}-g_{t\phi}g_{\phi\phi}g_{\theta\theta}\left(\frac{d\delta}{dt}\right)^2-\tilde{\zeta}^{1/2}}{g_{tt}g_{t\phi}^2-g_{t\phi}^2g_{\theta\theta}\left(\frac{d\delta}{dt}\right)^2-g_{tt}^2g_{\phi\phi}},
\ee
with
\ba
\tilde{\zeta}&=&g_{t\phi}^6-3g_{tt}g_{t\phi}^4g_{\phi\phi}-g_{t\phi}^4g_{\phi\phi}g_{\theta\theta}\left(\frac{d\delta}{dt}\right)^2+3g_{tt}^2g_{t\phi}^2g_{\phi\phi}^2+2g_{tt}g_{t\phi}^2g_{\phi\phi}^2g_{\theta\theta}\left(\frac{d\delta}{dt}\right)^2-g_{tt}^3g_{\phi\phi}^3\nn\\
&-&g_{tt}^2g_{\phi\phi}^3g_{\theta\theta}\left(\frac{d\delta}{dt}\right)^2.
\ea
$\tilde{\eta}$ is evaluated at $r = b$.

By solving Eq.~\eqref{eqn:b4} for $\frac{dt}{dr}$ and expanding the solution around $\theta=\pi/2$, we find, to first order in expansion parameter $\delta$, the same expression as Eq.~\eqref{eqn:Pi}. Therefore, we conclude that the (differential) time delay does not change if the trajectory of light deviate from the equatorial plane of the rotating black hole by a small angle.

\section{Reversing the spin}\label{app:rev}

In Sec.~\ref{sec:ps} we have studied the direct images in the case for which the direction of the spin of black hole is in the same sense as the direction of primary light ray rotation. Here we reverse the black hole spin and assume that the black hole is rotating in the same direction as the momentum of the light ray producing the secondary image.

\begingroup
\begin{table*}
	\caption{
		Image positions $\theta$ and deflection angles $\hat{\alpha}$ of primary and secondary images due to lensing by rotating (${\rm ro}^-$) as well as non-rotating (nr) black holes. Here the light coming from the primary image opposes the sense of rotation of the black hole. $\beta$ is the source angular position and the subscripts $p$ and $s$ refer to primary and secondary images, respectively. All angles are in {\em arcseconds}. We have taken $m$, $D_d$, $\mathcal{D}$, and $a^*$ as in Table \ref{tab:ps_ang}.
	}\label{tab:ps_rev_ang}
	\begin{ruledtabular}
		\begin{tabular}{l cccc cccc}
			\multicolumn{1}{c}{$\beta$}&
			\multicolumn{4}{c}{Rotating black hole}&
			\multicolumn{4}{c}{Non-rotating black hole}\\
			&$\theta_{p,{\rm ro}^-} $&$\hat{\alpha}_{p,{\rm ro}^-}$&$\theta_{s,{\rm ro}^-}$&$\hat{\alpha}_{s,{\rm ro}^-}$&$\theta_{p,{\rm nr}}$&$\hat{\alpha}_{p,{\rm nr}}$&$\theta_{s,{\rm nr}}$&$\hat{\alpha}_{s,{\rm nr}}$\\
			\hline
			$0	 	 $&$  1.4462845   $&$	2.8925690    		$&$  1.4462896  $&$    	  2.8925792     $&$  1.4462850     $&$	  2.8925701   	  $&$ 1.4462850    $&$     2.8925701        $\\
			$1	 	 $&$  1.4467818	  $&$   2.8915725    		$&$  1.4457885  $&$	   	  2.8935770     $&$  1.4467857	   $&$    2.8915713   	  $&$ 1.4457873	   $&$	   2.8935745        $\\
			$2	 	 $&$  1.4472814	  $&$   2.8905723    		$&$  1.4452892  $&$		  2.8945784     $&$  1.4472863	   $&$    2.8905726   	  $&$ 1.4452875    $&$	   2.8945750        $\\
			$3	 	 $&$  1.4477861	  $&$   2.8895721    		$&$  1.4447899  $&$    	  2.8955798     $&$  1.4477869	   $&$    2.8895739   	  $&$ 1.4447878    $&$	   2.8955755        $\\
			$4	 	 $&$  1.4482859	  $&$   2.8885719    		$&$  1.4442906  $&$    	  2.8965812     $&$  1.4482876	   $&$    2.8885752   	  $&$ 1.4442880    $&$     2.8965760        $\\
			$5	 	 $&$  1.4487858	  $&$   2.8875717    		$&$  1.4437913  $&$		  2.8975826     $&$  1.4487882	   $&$    2.8875764   	  $&$ 1.4437883    $&$	   2.8975765        $\\
			$6	 	 $&$  1.4492857	  $&$   2.8865715   		$&$  1.4432920  $&$    	  2.8985841     $&$  1.4492889	   $&$    2.8865777  	  $&$ 1.4432885	   $&$     2.8985771        $\\
		\end{tabular}
	\end{ruledtabular}
\end{table*}
\endgroup

\begingroup
\begin{table*}
	\caption{
		Magnification $\mu$ and (differential) time delay $\tau$ ($t_d = \tau_s-\tau_p$) of primary and secondary images due to lensing by rotating (${\rm ro}^-$) as well as non-rotating (nr) black holes. Here the light coming from the primary image opposes the sense of rotation of the black hole. $\beta$ and the subscripts $p$ and $s$ are as in Table \ref{tab:ps_ang}. (Differential) time delays are in {\em seconds}. $m$, $D_d$, $\mathcal{D}$, and $a^*$ are as in Table \ref{tab:ps_ang}.
	}\label{tab:ps_rev_time}
	\begin{ruledtabular}
		\begin{tabular}{l cccc cccc}
			\multicolumn{1}{c}{$\beta$}&
			\multicolumn{4}{c}{Rotating black hole}&
			\multicolumn{4}{c}{Non-rotating black hole}\\
			&$\mu_{p,{\rm ro}^-} $&$\tau_{p,{\rm ro}^-}$&$\mu_{s,{\rm ro}^-}$&$t_{d,{\rm ro}^-}$&$\mu_{p,{\rm nr}}$&$\tau_{p,{\rm nr}}$&$\mu_{s,{\rm nr}}$&$t_{d,{\rm nr}}$\\
			\hline
			$0	 	 $&$  \times  	  $&$	988.82368   	$&$  \times	     $&$    	  0.0005390     $&$  \times    	   $&$	  988.82417 	  $&$ \times  	    $&$    0      		    $\\
			$1		 $&$  723.24185	  $&$   988.76907   	$&$  -721.88022  $&$	   	  0.1097064     $&$  724.31554	   $&$    988.76968 	  $&$ -722.52663    $&$	   0.1086857        $\\
			$2		 $&$  361.74587	  $&$   988.71471 		$&$  -360.81546  $&$		  0.2184778     $&$  362.28309	   $&$    988.71524 	  $&$ -361.13844    $&$	   0.2175866        $\\
			$3	 	 $&$  241.24721	  $&$   988.66039 		$&$  -240.46054  $&$    	  0.3272492     $&$  241.60560	   $&$    988.66084 	  $&$ -240.67571    $&$	   0.3264876        $\\
			$4	 	 $&$  180.99788	  $&$   988.60610 		$&$  -180.28308  $&$    	  0.4360207     $&$  181.26686	   $&$    988.60647 	  $&$ -180.44435    $&$    0.4353886        $\\	
			$5	 	 $&$  144.84829	  $&$   988.55185 		$&$  -144.17660  $&$		  0.5447922     $&$  145.06362	   $&$    988.55214 	  $&$ -144.30553    $&$	   0.5442896        $\\
			$6	 	 $&$  120.74855	  $&$   988.49764 		$&$  -120.10562  $&$    	  0.6535638     $&$  120.92812	   $&$    988.49785 	  $&$ -120.21298    $&$    0.6531907        $\\
		\end{tabular}
	\end{ruledtabular}
\end{table*}
\endgroup

In Tables \ref{tab:ps_rev_ang} and \ref{tab:ps_rev_time} we have presented the lensing parameters of primary and secondary images for the case in which the light ray coming from the primary image opposes the sense of black hole spin, and compare them with the parameters of static black hole lensing. We have taken the black hole to be halfway between the source and observer. Similar to what we observed in Sec.~\ref{sec:ps}, the angular position of primary image increases and that of secondary image decreases by increasing the source angular position $\beta$. Also, the deflection angle associated with the primary (secondary) image decreases (increases) by increasing $\beta$. From Table \ref{tab:ps_rev_time} we see that the time delay of primary image decreases and the differential time delay increases by increasing $\beta$.

Comparing the values in Tables \ref{tab:ps_rev_ang} and \ref{tab:ps_rev_time} with those presented in Tables \ref{tab:ps_ang} and \ref{tab:ps_time} we see that for $\beta = 0$, the lensing parameters of primary image in Tables \ref{tab:ps_rev_ang} and \ref{tab:ps_rev_time} has been changed with the parameters of secondary image in Tables \ref{tab:ps_ang} and \ref{tab:ps_time} and vise versa. From Table \ref{tab:ps_rev_ang} we find that the angular position of primary image is larger for static black hole and that of secondary image is larger for rotating black hole. This is in contrast to what we have observed in Table \ref{tab:ps_ang} where the angular position of primary (secondary) image is larger rotating (static) black hole. Also, here the deflection angle of primary (secondary) image is larger for static (rotating) black hole.

In Table \ref{tab:ps_rev_time} we see that the time delay of primary image is larger for static black hole. This is also in contrast to what we have observed in Sec. \ref{sec:ps} where the time delay of primary image is larger for rotating black hole. On the other hand, in Table \ref{tab:ps_rev_time} we see that the differential time delay is larger for rotating black hole, while in Table \ref{tab:ps_time} we saw that the differential time delay is larger for static black hole. A similar observation can be made for magnifications.

Now, how we can find the direction (and the magnitude) of the black hole spin? As   mentioned in the main text, in standard lensing  $\beta$ cannot be measured directly. However, we can find it from the angular positions of the images; the argument we made in Sec.~\ref{sec:ps} can be generalized to the more realistic case in which the axis of rotation is not known. In fact one of the relations $\beta_{p,{\rm ro}^-}=\beta_{s,{\rm ro}^-}$, $\beta_{p,{\rm ro}}=\beta_{s,{\rm ro}}$, or $\beta_{p,{\rm nr}}=\beta_{s,{\rm nr}}$ should be true. This way we can find $\beta$ and find out if the black hole is rotating and, if so, in which direction.

\section{Dependence of first order relativistic images on $\mathcal{D}$}

\begin{figure}[htp]
	\centering
	\includegraphics[width=0.48\textwidth]{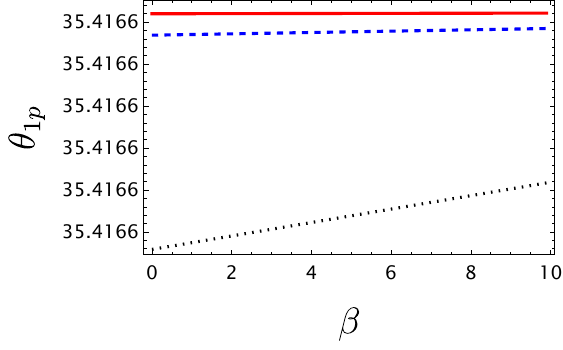}
	\includegraphics[width=0.48\textwidth]{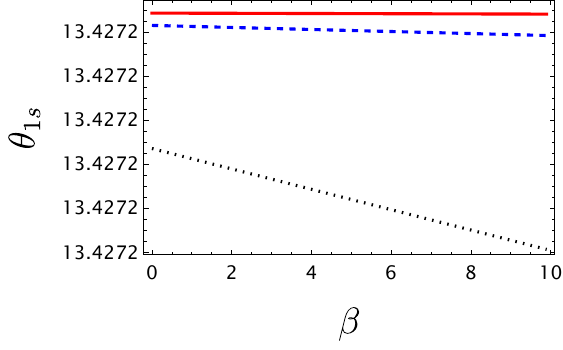}
	\caption{\textit{Left}: Angular position of first order relativistic images on the same side as the primary image. \textit{Right}: Angular position of first order relativistic images on the same side as the  secondary image. Both angles are in {\em microarcseconds}. $m$, $D_d$, and $a$ are as in Table \ref{tab:ps_ang}. The solid red curves correspond to $\mathcal{D}=0.5$, the dashed blue curves to $\mathcal{D}=0.05$, and the dotted black curves to  $\mathcal{D}=0.005$.
	}
	\label{fig:1st_theta}
\end{figure}

\begin{figure}[htp]
	\centering
	\includegraphics[width=0.48\textwidth]{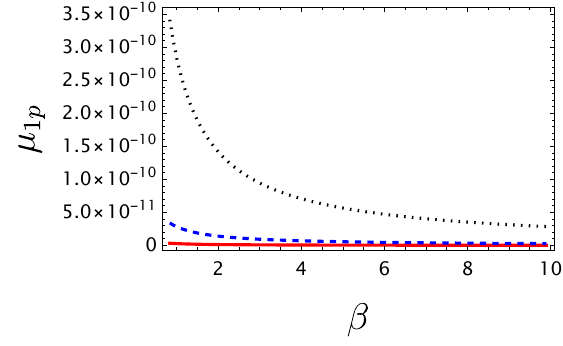}
	\includegraphics[width=0.48\textwidth]{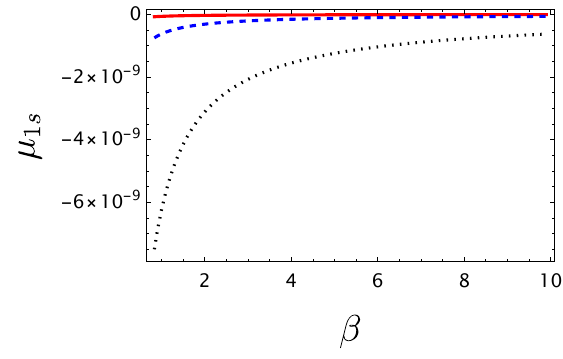}
	\caption{\textit{Left}: Magnification of first order relativistic images on the same side as the primary image. \textit{Right}: Magnification of first order relativistic images on the same side as the  secondary image. $\beta$ is in {\em microarcseconds}. $m$, $D_d$, and $a$ are as in Table \ref{tab:ps_ang}. 
		The solid red curves correspond to $\mathcal{D}=0.5$, the dashed blue curves to $\mathcal{D}=0.05$, and the dotted black curves to  $\mathcal{D}=0.005$.
	}
	\label{fig:1st_mu}
\end{figure}

\begin{figure}[htp]
	\centering
	\includegraphics[width=0.5\textwidth]{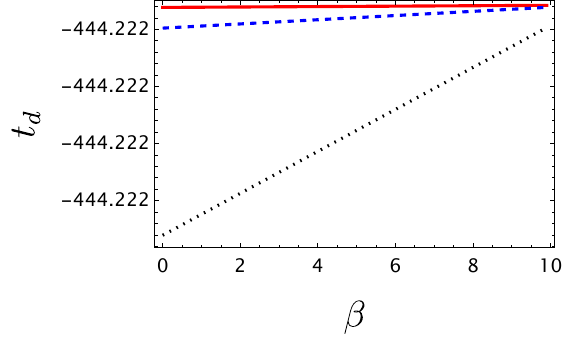}
	\caption{Differential time delay associated with the first order relativistic images. $\beta$ is in {\em microarcseconds} and differential time delays are in {\em seconds}. $m$, $D_d$, and $a$ are as in Table \ref{tab:ps_ang}. The solid red curves correspond to $\mathcal{D}=0.5$, the dashed blue curves to $\mathcal{D}=0.05$, and the dotted black curves to  $\mathcal{D}=0.005$.}
	\label{fig:1st_time}
\end{figure}

In Sec.~\ref{sec:rel} we have considered first order relativistic images for~$\mathcal{D}=0.5$, in which the black hole is halfway between the source and observer. In Figs.~\ref{fig:1st_theta}, \ref{fig:1st_mu}, and \ref{fig:1st_time} we illustrate how various lensing parameters of first order relativistic images depend on $\mathcal{D}$.

In Figure \ref{fig:1st_theta}, we see that for $\mathcal{D}=0.5$ the angular position of the first order relativistic image on the same side as the primary and secondary images are nearly insensitive to the value of $\beta$. For smaller values of $\mathcal{D}$, shown by the blue dashed ($\mathcal{D}=0.05$) and black dotted ($\mathcal{D}=0.005$) curves, we see that $\theta_{1p}$ increases and $\theta_{1s}$ decreases with increasing $\beta$. It is also obvious that for a fixed value of $\beta$ both $\theta_{1p}$ and $\theta_{1s}$ decrease as  $\mathcal{D}$ decreases.

In Figure \ref{fig:1st_mu} we plot the magnification $\mu$ of first order relativistic images produced by a rotating black hole as a function of $\beta$. We see that the magnification of a first order image on primary side and the absolute value of the magnification of a first order image on secondary side both decrease for increasing $\beta$. However for any given $\beta$ both    increase with decreasing $\mathcal{D}$.  In Figure \ref{fig:1st_time}, we plot the differential time delay $t_d$ of the first order relativistic images; we see that this quantity increases for increasing $\beta$, also for smaller values of $\mathcal{D}$, $t_d$ is smaller.

\section{Second order relativistic images}\label{app:2nd}

\begingroup
\begin{table}
	\scriptsize
	\caption{
		Image positions $\theta$ and deflection angle $\hat{\alpha}$ of second order relativistic images due to lensing by rotating (ro) as well as non-rotating (nr) black holes. The subscripts $2p$ and $2s$ refer to the second order relativistic images produced on the same side as the primary and secondary images, respectively. All angles are in {\em microarcseconds}. $m$, $D_d$, $\mathcal{D}$, and $a$ are as in Table \ref{tab:ps_ang}. The values are the same for all $\beta$ of the order of microarcseconds.
	}\label{tab:2nd_ang}
	\begin{ruledtabular}
		\begin{tabular}{l cccc cccc}
			\multicolumn{1}{c}{$\beta$}&
			\multicolumn{4}{c}{Rotating black hole}&
			\multicolumn{4}{c}{Non-rotating black hole}\\
			&$\theta_{2p,{\rm ro}} $&$\hat{\alpha}_{2p,{\rm ro}}$&$\theta_{2s,{\rm ro}}$&$\hat{\alpha}_{2s,{\rm ro}}$&$\theta_{2p,{\rm nr}}$&$\hat{\alpha}_{2p,{\rm nr}}$&$\theta_{2s,{\rm nr}}$&$\hat{\alpha}_{2s,{\rm nr}}$\\
			\hline
			$0	 	 $&$  35.410209   $&$2.5920024 \times 10^{12}$&$ 12.017557 $&$2.5920024 \times 10^{12}$&$  26.348094       $&$2.5920024 \times 10^{12}$&$26.348094$&$2.5920024 \times 10^{12}$\\
		\end{tabular}
	\end{ruledtabular}
\end{table}
\endgroup

\begingroup
\begin{table*}
	\caption{
		Magnifications $\mu$, time delay $\tau$, and differential time delay $t_d=\tau_{2s}-\tau_{2p}$ of second order relativistic images due to lensing by rotating (ro) as well as non-rotating (nr) black holes. The subscripts $2p$ and $2s$ refer to the second order relativistic images produced on the same side as the primary and secondary images, respectively. $\beta$ is in {\em microarcseconds} and (differential) time delays are in {\em seconds}. $m$, $D_d$, $\mathcal{D}$, and $a$ are as in Table \ref{tab:ps_ang}.
	}\label{tab:2nd_time}
	\begin{ruledtabular}
		\begin{tabular}{l cccc cccc}
			\multicolumn{1}{c}{$\beta$}&
			\multicolumn{4}{c}{Rotating black hole}&
			\multicolumn{4}{c}{Non-rotating black hole}\\
			&$\mu_{2p,{\rm ro}}$&$\tau_{2p,{\rm ro}}$&$\mu_{2s,{\rm ro}}$&$t_{d,{\rm ro}}$&$\mu_{2p,{\rm nr}}$&$\tau_{2p,{\rm nr}}$&$\mu_{2s,{\rm nr}}$&$t_{d,{\rm nr}}$\\
			\hline
			$0$&$	   \times    	    $&$  3605.845$&$  \times   			  $&$  -1001.889$&$	  	   \times   	     $&$ 3186.438    $&$  		\times     		   $&$     0     		$\\
			$1$&$ 8.2 \times 10^{-16}   $&$  3605.845$&$  -1.2 \times 10^{-11}$&$  -1001.889$&$    1.6 \times 10^{-14}   $&$ 3186.438    $&$ -1.6 \times 10^{-14}	   $&$1.97\times 10^{-9}$\\
			$2$&$ 4.1 \times 10^{-16}   $&$  3605.845$&$  -5.8 \times 10^{-12}$&$  -1001.889$&$    7.8 \times 10^{-15}   $&$ 3186.438    $&$ -7.8 \times 10^{-15}	   $&$3.94\times 10^{-9}$\\
			$3$&$ 2.7 \times 10^{-16}   $&$  3605.845$&$  -3.9 \times 10^{-12}$&$  -1001.889$&$    5.2 \times 10^{-15}   $&$ 3186.438    $&$ -5.2 \times 10^{-15}	   $&$5.91\times 10^{-9}$\\
			$4$&$ 2.0 \times 10^{-16}   $&$  3605.845$&$  -2.9 \times 10^{-12}$&$  -1001.889$&$    3.9 \times 10^{-15}   $&$ 3186.438    $&$ -3.9 \times 10^{-15}	   $&$7.88\times 10^{-9}$\\
			$5$&$ 1.6 \times 10^{-16}   $&$  3605.845$&$  -2.3 \times 10^{-12}$&$  -1001.889$&$    3.1 \times 10^{-15}   $&$ 3186.438    $&$ -3.1 \times 10^{-15}	   $&$9.85\times 10^{-9}$\\
			$6$&$ 1.4 \times 10^{-16}   $&$  3605.845$&$  -1.9 \times 10^{-12}$&$  -1001.889$&$    2.6 \times 10^{-15}   $&$ 3186.438    $&$ -2.6 \times 10^{-15}	   $&$1.18\times 10^{-8}$\\
		\end{tabular}
	\end{ruledtabular}
\end{table*}
\endgroup

In tables \ref{tab:2nd_ang} and \ref{tab:2nd_time} we present the lensing parameters of second order relativistic images for both the rotating and non-rotating cases. Compared to the first order relativistic images, these images are closer to the black hole and experience correspondingly greater demagnification. Just as for first order relativistic images, the angular position  and deflection angle of a second order relativistic image are nearly insensitive to the value of the  angular position of the source. We also note from Table \ref{tab:2nd_ang} that, for a non-rotating lens, the angular positions of second order images produced on the same side as the secondary image are nearly the same as those produced on the same side as the primary image, i.e.~$\theta_{2s,{\rm nr}}\simeq\theta_{2p,{\rm nr}}$. However  for a rotating lens the angular position of an image produced on the primary side is larger than the angular position of an image produced on the secondary side. In fact, the angular position of a second order image produced on the primary (secondary) side is larger (smaller) if the lens is rotating.

From Table \ref{tab:2nd_time} we see that the magnifications of second order relativistic images decrease by increasing the angular source position
if they are on  the same side as the primary  image, and their absolute values decrease if they are on the same side as the secondary  image.
For a static lens, $\mu_{2s,{\rm nr}}\simeq -\mu_{2p,{\rm nr}}$, whereas for a rotating lens, the absolute value of the magnification of images produced on the secondary side is about four orders of magnitude larger than that produced on the primary side. We find from Table \ref{tab:2nd_time} that the magnification of second order images produced on the primary side is smaller if the lens is rotating. However, the absolute value of the magnification of images produced on secondary side is larger for the rotating black hole.

The time delay of a second order relativistic image on the primary side is larger than that of its first order relativistic counterpart  for both a rotating and a static lens. We also note that, for a rotating lens, the absolute values of differential time delays of second order relativistic images are larger than those of first order relativistic images. In Table \ref{tab:2nd_time} we also see that the time delays associated with the second order relativistic images produced on the same side as the primary image are nearly insensitive to the value of $\beta$ --- to be more accurate, they decrease very slightly by increasing $\beta$. This time delay is larger if the black hole is rotating. Of more importance is the differential time delay. For static black holes, this quantity increases by increasing the angular source position. However, for rotating black hole, the absolute value of the differential time delay decreases  slightly by increasing $\beta$.

The dependence of the lensing parameters of the second order relativistic images on $\mathcal{D}$ are similar to those of the first order relativistic images, and so we do not present plots of their image position, magnification, and differential time delay; they are qualitatively the same as those depicted in Figs.~\ref{fig:1st_theta}, \ref{fig:1st_mu}, and \ref{fig:1st_time}. For decreasing $\mathcal{D}$ the angular positions of second order relativistic images decreases, the separation of $\theta_{2s}$ and $\theta_{2p}$ increases, the absolute value of the magnification $\mu$ becomes larger. Differential time delays of second order relativistic images are also  insensitive to $\mathcal{D}$.

\section{Retrolensing}\label{app:retro}

\begin{figure}[htp]
	\centering
	\includegraphics[width=0.48\textwidth]{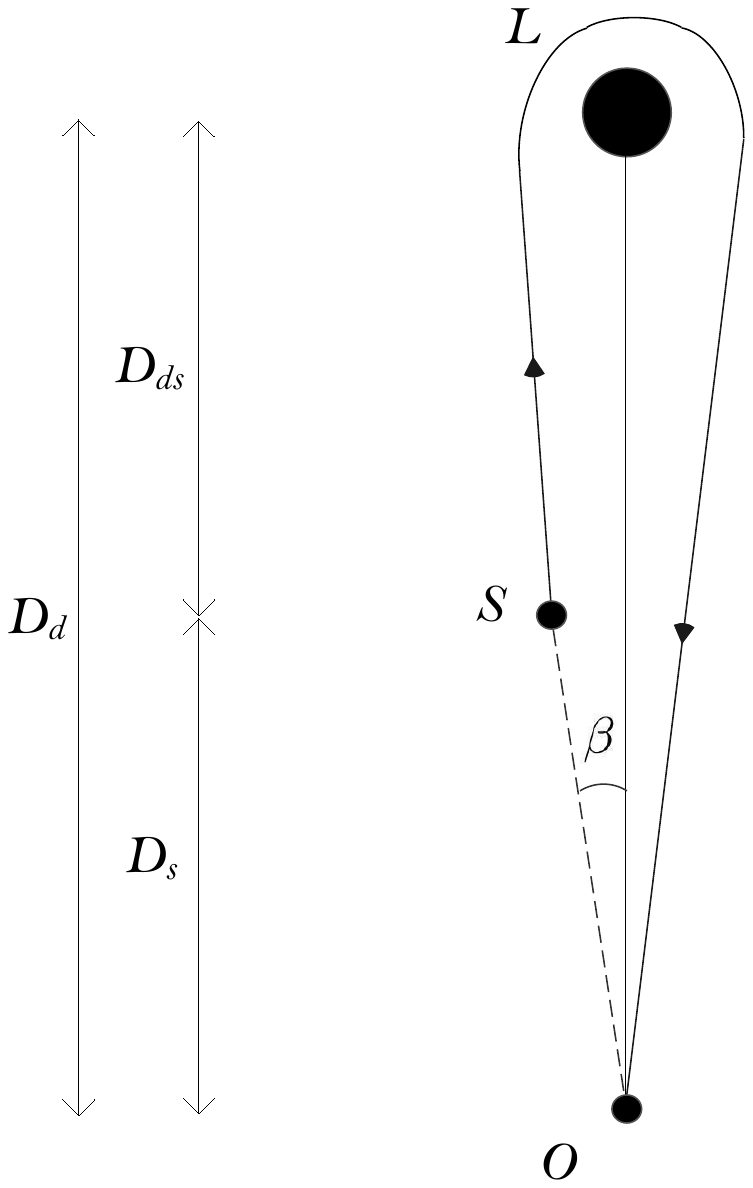}
	\includegraphics[width=0.48\textwidth]{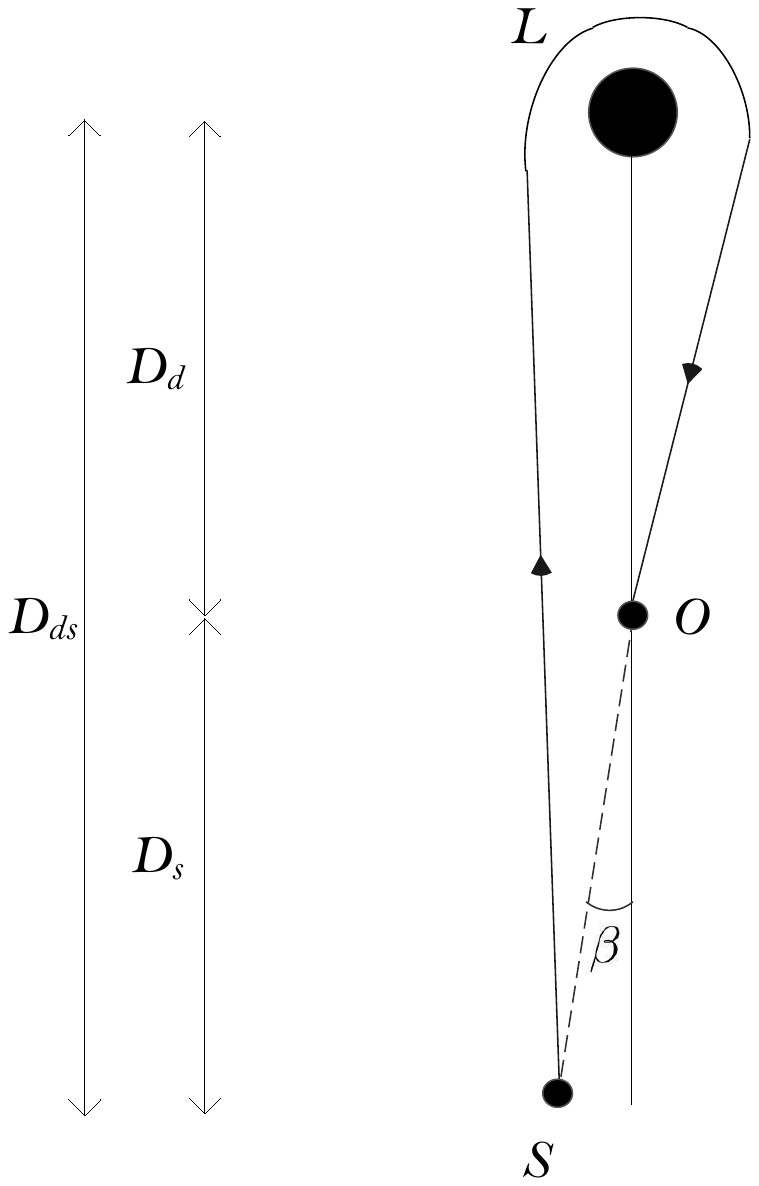}
	\caption{\textit{Left}: Retrolensing diagram in which the source is situated between the observer and the lens. \textit{Right}: Retrolensing diagram in which the observer is situated between the source and the lens.
	}
	\label{fig:retro_diag}
\end{figure}

So far we have investigated standard lensing in which the lens (black hole) is situated between source and the observer.
Other source-lens-observer orders are possible, however, as shown  in Figure \ref{fig:retro_diag}. These situations are referred to as retrolensing and were first studied in \cite{Holz:2002uf}.

The lens equation in retrolensing is similar to Virbhadra-Ellis lens equation \eqref{eqn:lens_eq}
\be
\tan\beta=\pm\left\{\tan\theta-c_3\left[\tan\theta+\tan\left(\hat{\alpha}-\theta\right)\right]\right\},
\ee
where $c_3=\frac{D_s}{D_d}$ for the retrolensing case in which the source is between the lens and the observer and $c_3=\frac{D_s}{D_{ds}}$ for the case in which the observer is situated between the source and the lens \cite{Eiroa:2004gh}. Here again the plus sign is for the primary image and the minus sign is for the secondary image. However, in retrolensing the primary image is on the opposite side of the line of sight compared to the source and the secondary image is on the same side as the source.

The relations that we have presented in Sec.~\ref{sec:lensing} for the deflection angle and magnification of images also work for retrolensing. However, the relation for time delay should be modified. Let us first consider the case in which the source is situated between the observer and the lens. As we see in the left panel of Figure \ref{fig:retro_diag}, in this case $D_d=D_s+D_{ds}$. Therefore
\be
D_{ds}=\left(1-c_3\right)D_d.
\ee
The time delay can be found by
\be
\tau(b)=\int_{b}^{r_s}\Pi\,dr+\int_{b}^{D_d}\Pi\,dr-c_3 D_d\sec\beta,
\ee
with the auxiliary function $\Pi$ given by Eq.~\eqref{eqn:Pi} and
\be
r_s=\sqrt{\left(1-c_3\right)^2 D_d^2 +c_3^2D_d^2\tan^2\beta}.
\ee

Now consider the retrolensing case in which the observer is situated between the source and the black hole. From the right panel of Figure \ref{fig:retro_diag} we see $D_{ds}=D_d+D_s$. Thus we find
\ba
D_s&=&\frac{c_3 D_d}{1-c_3},\\
D_{ds}&=&\frac{D_d}{1-c_3}.
\ea
We define the time delay by
\be
\tau(b)=\int_{b}^{r_s}\Pi\,dr+\int_{b}^{D_d}\Pi\,dr-\frac{c_3}{1-c_3} D_d\sec\beta,
\ee
in which
\be
r_s=\sqrt{\left(\frac{D_d}{1-c_3}\right)^2+\left(\frac{c_3 D_d}{1-c_3}\tan\beta\right)^2}.
\ee

In what follows we only consider the first case of retrolensing in which the source is situated between the observer and the black hole. In Tables \ref{tab:retro_ang} and \ref{tab:retro_time} we have presented the lensing parameters for a configuration where  the source is halfway between the observer and the black hole, i.e.~$c_3=0.5$.  In Table \ref{tab:retro_ang} we can see that the deflection angle of a primary image is larger if the lens is rotating but for a secondary image it is larger if the lens is static. The deflection angle of  the primary image decreases as $\beta$ increases, whereas the deflection angle of secondary image increases.

\begingroup
\begin{table}
	\scriptsize
	\caption{
		Image positions $\theta$ and deflection angle $\hat{\alpha}$ of images produced in retrolensing by rotating (ro) as well as non-rotating (nr) black holes. The subscripts $s$ and $p$ refer to the retrolensing images produced on the same side as the source and on the opposite site, respectively. $\beta$ and $\theta$ are in {\em microarcseconds} and $\hat{\alpha}$ is in radians. $m$, $D_d$, $\mathcal{D}$, and $a$ are as in Table \ref{tab:ps_ang}. We have taken $c_3=0.5$.
	}\label{tab:retro_ang}
	\begin{ruledtabular}
		\begin{tabular}{l cccc cccc}
			\multicolumn{1}{c}{$\beta$}&
			\multicolumn{4}{c}{Rotating black hole}&
			\multicolumn{4}{c}{Non-rotating black hole}\\
			&$\theta_{p,{\rm ro}} $&$\hat{\alpha}_{p,{\rm ro}}-\pi$&$\theta_{s,{\rm ro}}$&$\hat{\alpha}_{s,{\rm ro}}-\pi$&$\theta_{p,{\rm nr}}$&$\hat{\alpha}_{p,{\rm nr}}-\pi$&$\theta_{s,{\rm nr}}$&$\hat{\alpha}_{s,{\rm nr}}-\pi$\\
			\hline
			$0	 	 $&$  35.796060   $&$3.470850 \times 10^{-10}$&$ 16.530462 $&$1.602766 \times 10^{-10}$&$27.163430$&$2.633879 \times 10^{-10}$&$27.163430$&$2.633879 \times 10^{-10}$\\
			$1		 $&$  35.796060	  $&$3.373888 \times 10^{-10}$&$ 16.530462 $&$1.699728 \times 10^{-10}$&$27.163430$&$2.536916 \times 10^{-10}$&$27.163430$&$2.730842 \times 10^{-10}$\\
			$2		 $&$  35.796060	  $&$3.276925 \times 10^{-10}$&$ 16.530462 $&$1.796691 \times 10^{-10}$&$27.163430$&$2.439954 \times 10^{-10}$&$27.163430$&$2.827804 \times 10^{-10}$\\
			$3	 	 $&$  35.796060	  $&$3.179962 \times 10^{-10}$&$ 16.530462 $&$1.893654 \times 10^{-10}$&$27.163430$&$2.342991 \times 10^{-10}$&$27.163430$&$2.924767 \times 10^{-10}$\\
			$4	 	 $&$  35.796060	  $&$3.082999 \times 10^{-10}$&$ 16.530462 $&$1.990616 \times 10^{-10}$&$27.163430$&$2.246028 \times 10^{-10}$&$27.163430$&$3.021730 \times 10^{-10}$\\
			$5	 	 $&$  35.796060	  $&$2.986037 \times 10^{-10}$&$ 16.530462 $&$2.087579 \times 10^{-10}$&$27.163430$&$2.149066 \times 10^{-10}$&$27.163430$&$3.118692 \times 10^{-10}$\\
			$6	 	 $&$  35.796060	  $&$2.889074 \times 10^{-10}$&$ 16.530462 $&$2.184542 \times 10^{-10}$&$27.163430$&$2.052103 \times 10^{-10}$&$27.163430$&$3.215655 \times 10^{-10}$\\
		\end{tabular}
	\end{ruledtabular}
\end{table}
\endgroup

\begingroup
\begin{table}
	\scriptsize
	\caption{
		Magnifications $\mu$, time delay $\tau$, and differential time delay $t_d=\tau_{1s}-\tau_{1p}$ of images produced in retrolensing by rotating (ro) as well as non-rotating (nr) black holes. The subscripts $s$ and $p$ refer to the retrolensing images produced on the same side as the source and on the opposite site, respectively. $\beta$ is in {\em microarcseconds} and (differential) time delays are in {\em seconds}. $m$, $D_d$, $\mathcal{D}$, and $a$ are as in Table \ref{tab:ps_ang}. We have taken $c_3=0.5$.
	}\label{tab:retro_time}
	\begin{ruledtabular}
		\begin{tabular}{l cccc cccc}
			\multicolumn{1}{c}{$\beta$}&
			\multicolumn{4}{c}{Rotating black hole}&
			\multicolumn{4}{c}{Non-rotating black hole}\\
			&$\mu_{p,{\rm ro}}$&$\tau_{p,{\rm ro}}$&$\mu_{s,{\rm ro}}$&$t_{d,{\rm ro}}$&$\mu_{p,{\rm nr}}$&$\tau_{p,{\rm nr}}$&$\mu_{s,{\rm nr}}$&$t_{d,{\rm nr}}$\\
			\hline
			$0$&$	   \times    	$&$  8.005538 \times 10^{11}$&$       \times   		 $&$-189.533$&$  	   \times   	 $&$8.005538 \times 10^{11}$&$  		\times     $&$     0     		$\\
			$1$&$1.8 \times 10^{-10}$&$  8.005538 \times 10^{11}$&$  -3.2 \times 10^{-10}$&$-189.533$&$2.3 \times 10^{-10}   $&$8.005538 \times 10^{11}$&$ -2.3 \times 10^{-10}$&$2.04\times 10^{-9}$\\
			$2$&$8.9 \times 10^{-11}$&$  8.005538 \times 10^{11}$&$  -1.6 \times 10^{-10}$&$-189.533$&$1.1 \times 10^{-10}   $&$8.005538 \times 10^{11}$&$ -1.1 \times 10^{-10}$&$4.09\times 10^{-9}$\\
			$3$&$5.9 \times 10^{-11}$&$  8.005538 \times 10^{11}$&$  -1.1 \times 10^{-10}$&$-189.533$&$7.6 \times 10^{-11}   $&$8.005538 \times 10^{11}$&$ -7.6 \times 10^{-11}$&$6.13\times 10^{-9}$\\
			$4$&$4.5 \times 10^{-11}$&$  8.005538 \times 10^{11}$&$  -8.1 \times 10^{-11}$&$-189.533$&$5.7 \times 10^{-11}   $&$8.005538 \times 10^{11}$&$ -5.7 \times 10^{-11}$&$8.18\times 10^{-9}$\\
			$5$&$3.6 \times 10^{-11}$&$  8.005538 \times 10^{11}$&$  -6.4 \times 10^{-11}$&$-189.533$&$4.6 \times 10^{-11}   $&$8.005538 \times 10^{11}$&$ -4.6 \times 10^{-11}$&$1.02\times 10^{-8}$\\
			$6$&$3.0 \times 10^{-11}$&$  8.005538 \times 10^{11}$&$  -5.4 \times 10^{-11}$&$-189.533$&$3.8 \times 10^{-11}   $&$8.005538 \times 10^{11}$&$ -3.8 \times 10^{-11}$&$1.23\times 10^{-8}$\\
		\end{tabular}
	\end{ruledtabular}
\end{table}
\endgroup

We also see in Table \ref{tab:retro_ang} that the angular positions of the primary and secondary images are nearly the same for a non-rotating lens, whereas they are very different if the lens is rotating.  The image position is nearly insensitive to the angular source position.  For primary images, the angular position is larger if the lens is rotating, but for secondary images the angular position is larger if the lens is static.  In Table \ref{tab:retro_time} we observe that, like relativistic images, the retrolensing images are highly
demagnified\footnote{We note that in previous work we erroneously referred to retrolensing images as  first order relativistic images \cite{Ashoorioon:2021gjs}. The actual first order relativistic images are what we called  second order relativistic images.}.
For static black holes the absolute value of the magnification of a secondary image is almost equal to the magnification of a corresponding primary image. However, for a rotating black hole, the absolute value of the magnification of a secondary image is slightly larger than the magnification of a primary image. For primary images, the magnification is larger if the lens is static, but  for secondary images it is larger if the lens is rotating. Note also that the absolute value of the magnification decreases with increasing $\beta$.

Though not obvious from Table \ref{tab:retro_time}, the time delay of a primary image is larger for a rotating black hole as compared to a static black hole, and both time delays decrease with increasing $\beta$. The differential time delay for the rotating case is a large negative number  $\sim 10^{3}$ seconds, in contrast to   the differential time delay of   static black hole retrolensing, which is very small ($\sim 10^{-9}$ seconds). The retrolensing differential time delay of  static black holes slightly increases with $\beta$. However the absolute value of the differential time delay for the rotating black hole case decreases with increasing $\beta$.

Since in retrolensing $\beta$ can be measured directly, and the magnification in retrolensing is about one order of magnitude larger than that in relativistic images, we find that retrolensing provides a better probe to measure the black hole spin, compared to relativistic images.

\bibliography{mybib}

\end{document}